\documentclass[]{siamart171218}

\usepackage{lipsum}
\usepackage{amsfonts}
\usepackage{graphicx}
\usepackage{epstopdf}
\usepackage{algorithm}
\usepackage{enumerate, amsmath,amssymb, graphicx, multicol, array, mathrsfs, comment, dsfont, esint, mathtools, subfigure, url, float, multicol}
\ifpdf
  \DeclareGraphicsExtensions{.eps,.pdf,.png,.jpg}
\else
  \DeclareGraphicsExtensions{.eps}
\fi
\usepackage[compatible]{algpseudocode} 
\algnewcommand\algorithmicreturn{\textbf{return}}
\algnewcommand\RETURN{\State \algorithmicreturn}
\renewcommand{\algorithmiccomment}[1]{\bgroup\hfill//~#1\egroup}

\newcommand{\R}{\mathbb{R}}

\newcommand{\e}{\epsilon}

\newcommand{\B}[1]{\mathbf{#1}}

\headers{QUBO Algorithm for Eigenvector Problems}{B. Krakoff, S. Mniszewski, C. Negre}

\title{A QUBO Algorithm to Compute Eigenvectors of Symmetric Matrices \thanks{This article was submitted to the arXiv \today \funding{This research was supported in part by an
appointment with the National Science Foundation (NSF) Mathematical Sciences Graduate Internship (MSGI) Program
sponsored by the NSF Division of Mathematical Sciences. This program is administered by the Oak Ridge Institute for Science
and Education (ORISE) through an interagency agreement between the U.S. Department of Energy (DOE) and NSF. ORISE is
managed for DOE by ORAU. All opinions expressed in this paper are the author's and do not necessarily reflect the policies and
views of NSF, ORAU/ORISE, or DOE. We acknowledge LANL computational facilities.  
This document has been assigned the following Unlimited Release Number (LAUR): 'LA-UR-21-21969' dated Feb 26th, 2021.}}}

\author{Benjamin Krakoff \thanks{University of Michigan Mathematics Department 
(\email{bkrakoff@umich.edu}).}
\and Susan M. Mniszewski \thanks{Computer, Computational, \& Statistical Sciences Division, Los Alamos National Laboratory
  (\email{smm@lanl.gov}).}
  \and 
Christian F. A. Negre\thanks{Theoretical Division, Los Alamos National Laboratory 
  (\email{cnegre@lanl.gov}).}}

\usepackage{amsopn}

\makeatletter
\newcommand*{\addFileDependency}[1]{
  \typeout{(#1)}
  \@addtofilelist{#1}
  \IfFileExists{#1}{}{\typeout{No file #1.}}
}
\makeatother

\newcommand*{\myexternaldocument}[1]{%
    \externaldocument{#1}%
    \addFileDependency{#1.tex}%
    \addFileDependency{#1.aux}%
}

\ifpdf
\hypersetup{
  pdftitle={Controlled Precision QUBO-based Algorithm to Compute Eigenvectors of Symmetric Matrices},
  pdfauthor={Benjamin Krakoff, Susan M. Mniszewski, Christian F. A. Negre}
}
\fi

\myexternaldocument{ex_supplement}

\begin{document}
\maketitle
\begin{abstract}
  We describe an algorithm to compute the extremal eigenvalues and corresponding eigenvectors of a symmetric matrix by solving a sequence of Quadratic Binary Optimization problems. This algorithm is robust across many different classes of symmetric matrices, can compute the eigenvector/eigenvalue pair to essentially arbitrary precision, and with minor modifications can also solve the generalized eigenvalue problem. 
  Performance is analyzed on small random matrices and selected larger matrices from practical applications.
\end{abstract}



\section{Introduction}
The problem of computing eigenvectors and eigenvalues to a desired precision has many applications in science and mathematics, including web page ranking \cite{ilprints422}, planar embeddings \cite{Hall} and principal component analysis \cite{pearson_karl}, among many others. The recent development of new computing paradigms has led to the production of various kinds of \emph{annealers}, which are specialized hardware designed to solve Quadratic Binary Optimization Problems (QUBOs). Such annealers include D-Wave's quantumn annealers and Fujitsu's Digital Annealer, to name a few. This has lead to a corresponding interest in reformulating computational tasks as QUBOs and solving them using these annealers. This strategy has been applied to several problems including, graph partitioning \cite{10.1145/3149526.3149531}, solving polynomial equations \cite{Chang} and vertex coloring \cite{Kochenberger}. Here we compute eigenvectors of symmetric matrices by solving a sequence of QUBOs, which allow the eigenvectors and eigenvalues to be found to any desired precision. A mathematically similar approach to this problem is considered in \cite{Teplukhin_2020}, but accuracy is increased by increasing the size of the associated QUBO. In contrast, the proposed algorithm can compute eigenvectors to essentially arbitrary precision without increasing the size of the QUBOs, which can have as few as twice as many variables as the original eigenvalue problem. The trade-off for using small QUBOs is that more iterations are required. A nearly identical approach is considered in Appendix C of \cite{AQAE}, although here the effects of different parameters are more thoroughly studied, and the presentation gives a very general optimization framework. The performance data is collected using D-Wave's Ocean simulated annealing (SA) package.

The paper is organized as follows. The relevant mathematical background for symmetric matrices and use of QUBO solvers as a descent method is explained in \cref{sec:background}. The algorithm for computing the eigenvector/eigenvalue pair is given in \cref{sec:alg}. Experimental results with various parameters and matrices are presented in \cref{sec:results}, followed by the conclusion in \cref{sec:conclusion}.

\section{Methods}
\subsection{Mathematical Background}
\label{sec:background}
Let $A$ be a symmetric matrix. A well-known consequence of the spectral theorem is that the smallest eigenvalue $\lambda$ and corresponding eigenvector $\B{v}$ are global minima for the Rayleigh quotient $\B{x}^t A \B{x}/\B{x}^t\B{x}$ 
	\begin{align}
		\lambda = \min_{||\B{x}||=1} \B{x}^t A \B{x}, \ \ \B{v} = \underset{||\B{x}||=1}{\text{argmin }}\B{x}^t A \B{x}
		\label{rayleigh}
	\end{align}
	\indent The proposed algorithm uses a QUBO formulation of the problem to both obtain a good initial guess for the global minimum, and to implement an iterative descent from the initial guess. Similar to classical descent methods such as Newton Conjugate-Gradient and the BFGS algorithms \cite{Stoer}, the algorithm requires computing, but not inverting, a Hessian matrix at each descent step. We begin with an overview of QUBOs and how they can be used to approximately solve certain constrained quadratic optimization problems. \\
	\indent Let $\{0, 1\}^m$ denote the set of binary vectors of length $m$, and let $Q$ be a symmetric $m \times m$ matrix. The combinatorial optimization problem
	$$\underset{\B{x}_b \in \{0,1\}^m}{\text{argmin }} \B{x}_b^tQ \B{x}_b$$
	is called a \emph{quantum unconstrained binary optimization} problem, or QUBO, and it is known to be NP-hard \cite{1982JPhA...15.3241B}. Interest in casting various problems as QUBOs has increased due to the development of various kinds of annealers, which are a class of hardware that use ideas from statistical mechanics to produce approximate solutions to a QUBO. See for example \cite{DA} or \cite{Boixo}. \\
	\indent To solve a real-variable optimization problem using a QUBO, we require a method of approximating each real variable by $b$ binary variables. This number $b$ will be a parameter referred to as the number of bits. \\
	\indent Let's start with a few concrete examples of the arithmetic involved, beginning with a demonstration of how to multiply two real numbers $x = -.5$ and $y = .5$ using 2 bits. Form the \emph{precision vector} $\B{p} = (-1, .5)$ with corresponding \emph{precision matrix} $P = \B{p}^t\B{p} = \begin{pmatrix} 1 & -.5 \\ -.5 & .25 \end{pmatrix}$. Set $\B{x}_b = \begin{pmatrix} 1 \\ 1 \end{pmatrix}$ and $\B{y}_b = \begin{pmatrix} 0 \\ 1 \end{pmatrix}$ so that 
	\begin{align}
		& x = \B{p} \cdot \B{x}_b, \ \ y = \B{p} \cdot \B{y}_b \\
		& -.25 = x \cdot y = \B{x}_b^t \B{p}^t\B{p} \B{y}_b = (1, 1) \begin{pmatrix} 1 & -.5 \\ -.5 & .25 \end{pmatrix} \begin{pmatrix}
			0 \\ 1\end{pmatrix}
		\label{precexample}	
	\end{align}
	\indent Now let $Q$ be a symmetric matrix, and we shall demonstrate how to compute the quadratic form $(x, y, z)Q \begin{pmatrix}
		x \\ y \\ z
	\end{pmatrix}$ using binary variables. As above, set  $z = \B{p} \cdot \B{z}_b$ where $\B{z}_b$ is a binary vector and $z$ is the corresponding real number. We can rewrite the quadratic form as 
	\begin{align}
		(\B{x}_b^t, \B{y}_b^t, \B{z}_b^t)\begin{pmatrix}
			\B{p}^t & & \\
			& \B{p}^t & \\
			& & \B{p}^t
		\end{pmatrix} Q \begin{pmatrix}
		\B{p} & & \\
		& \B{p} & \\
		& & \B{p}
		\end{pmatrix} \begin{pmatrix}
		\B{x}_b \\ \B{y}_b \\ \B{z}_b
		\end{pmatrix}
		\label{tensorprod}
	\end{align}
	The middle three terms can be written more succinctly as $(I_3 \otimes p^t) Q (I_3 \otimes p) = Q \otimes P$ where $I_3$ is the $3 \times 3$ identity matrix and $\otimes$ is the tensor product.\\
	\indent Now we describe the construction in full generality. Given a \emph{precision vector} $\B{p} = (-1, \frac{1}{2}, \frac{1}{2^2}, \frac{1}{2^3}, \ldots \frac{1}{2^{b-1}})$ of length $b$, the set of integer multiples of $\frac{1}{2^{b-1}}$  in the interval $[-1, 1-\frac{1}{2^b}]$ is exactly the set
	\begin{align}
		C_{1, b} := \{\B{p} \cdot \B{x}_b | \B{x}_b \in \{0, 1\}^b \}
		\label{discsetVect}
	\end{align} 
	We use the sub-scripted $\B{x}_b$ as a convention to emphasize that $\B{x}_b$ is a binary vector, i.e. the subscript does not refer to the number of bits. More generally, the $n$-fold product of $C_{1, b}$ is the set
	\begin{align}
		C_{n, b} := \{(I_{n} \otimes \B{p}) \B{x}_b | \B{x}_b \in \{0, 1\}^{nb}\}
		\label{discsetMat}
	\end{align}
	where $I_n$ is the identity matrix, $\otimes$ is the tensor product. The set $C_{n, b}$ will be referred to as a discretized cube. Let $Q$ be a symmetric $n \times n$ matrix, $\B{r}$ an $n$-vector and suppose we want to solve the constrained quadratic programming problem
	\begin{align}
	\label{realquadprogram}
	\underset{\B{x} \in [-1,1]^n}{\text{argmin }} \B{r}^t\B{x} + \B{x}^tQ\B{x}
	\end{align}
	\indent To get an approximate solution to $\ref{realquadprogram}$ using a QUBO, first replace the unit cube by a discretized unit cube to get the optimization problem.
	\begin{align}
	\label{discreteApprox}
	\underset{\B{x} \in C_{n,b}}{\text{argmin }} \B{r}^t\B{x} + \B{x}^tQ\B{x}
	\end{align}
 Setting $m = nb$, $\B{x} = (I_n \otimes \B{p}) \B{x}_b$, and $P = \B{p}^t\B{p}$, this is equivalent to the QUBO
	\begin{align}
		\underset{\B{x}_b \in \{0, 1\}^m}{\text{argmin }} \ \B{r}^t(I_n \otimes \B{p}) \B{x}_b & + \B{x}_b^t(Q \otimes P)\B{x}_b \label{line1} \\
		& = \underset{\B{x}_b \in \{0, 1\}^m}{\text{argmin }} \ \B{x}_b^t \text{Diag}(\B{r}^t I_n \otimes \B{p}) \B{x}_b + \B{x}_b^t (Q \otimes P) \B{x}_b \label{line2}\\
		& = \underset{\B{x}_b \in \{0, 1\}^m}{\text{argmin }} \ \B{x}_b^t(\text{Diag}(\B{r}^t I_n \otimes \B{p}) + Q \otimes P) \B{x}_b \label{finalQUBO}
		\end{align}
	\indent Here $\text{Diag}(\B{v})$ refers to the diagonal matrix with entries from $\B{v}$. Going from lines \ref{line1} to \ref{line2} uses the following identity valid for binary vectors: $\B{v}^t\B{x}_b = \B{x}_b^t \text{Diag}(\B{v}) \B{x}_b$. \\
	\indent To summarize, given a number of bits $b$, this procedure approximates the real optimization problem in $n$ variables \ref{realquadprogram} with the QUBO \ref{finalQUBO} of size $n \cdot b$. We conclude by remarking that we are not restricted to the cube $[-1, 1]^n$. If we instead want to optimize over the cube $[-\delta, \delta]^n$, repeat the same construction with the precision vector $\delta \cdot \B{p}$.  \\
	\indent It is worth briefly discussing the error introduced by replacing the real cube with a discretized cube. Suppose we had an ideal annealer that always produces a best solution to \ref{discreteApprox}, call that solution $\B{x}_A$. Let $\B{x}_T$ be a best solution to \ref{realquadprogram}, and let $\hat{\B{x}_T}$ be the point in $C_{n, b}$ closest to $\B{x}_T$. The gradient of $\B{r}^t\B{x} + \B{x}^tQ\B{x}$ is $2Q\B{x} + \B{r}$, and $||\hat{\B{x}_T} - \B{x}_T|| \leq \sqrt{\frac{n}{2^b}}$, leading to the Lipschitz estimate
	\begin{align}
	\label{lipschitzest}
		|\B{r}^t\hat{\B{x}_T}+ \hat{\B{x}_T}^tQ\hat{\B{x}_T} - (\B{r}^t\B{x}_T + \B{x}_T^tQ\B{x}_T)| \leq \sqrt{\frac{n}{2^b}} \sup_{[-1, 1]^n} ||2Q\B{x} + \B{r}||
		\end{align}
	combining with the inequality $\B{r}^t\B{x}_T + \B{x}_T^tQ\B{x}_T \leq \B{r}^t\B{x}_A + \B{x}_A^tQ\B{x}_A \leq \B{r}^t\hat{\B{x}_T} + \hat{\B{x}_T}^tQ\hat{\B{x}_T}$ implies 
	\begin{align}
		|\B{r}^t\B{x}_A + \B{x}_A^tQ\B{x}_A - (\B{r}^t\B{x}_T + \B{x}_T^tQ\B{x}_T)| \leq \sqrt{\frac{n}{2^b}} \sup_{[-1, 1]^n} ||2Q\B{x} + \B{r}||
		\label{lipschitzest2}
	\end{align}
	This inequality makes a trade-off apparent. With more bits, the solution on the discretized cube will better approximate the true solution of \ref{realquadprogram} but will require solving a larger QUBO. Indeed, numerical experiments from subsequent sections will show that $b=2$ generally requires more iterations than $b=8$, indicating that the quality of the approximate solution at each step is worse, although interestingly using $b=2$ takes less time overall since solving smaller QUBOs is much faster. It is also worth noting that estimate \ref{lipschitzest} does not control the actual distance between solutions $||\B{x}_A - \B{x}_T||$. In the special case when $Q$ is positive definite, this distance can be controlled, but it would be interesting to have estimates in greater generality.\\
	\indent The annealers that one works with in practice are never ideal, and so will rarely return the absolute best solution $\B{x}_A$ but instead a response consisting of many samples $\B{x}_1, \B{x}_2, \ldots, \B{x}_l$ of good solutions with energies $E(\B{x}_i) \leq E(\B{x}_{i+1})$. (Here \emph{energy} of a solution $\B{x}_i$ refers to the value of the objective function at $\B{x}_i$). An obvious approach is to treat the lowest energy solution $\B{x}_0$ as the best approximation of $\B{x}_T$. A subtler approach that can reap great benefits in practice is to take a linear combination of the full response:
	\begin{equation}
	\frac{1}{l}\sum_{i=1}^l e^{-\beta(E(\B{x}_i) - E(\B{x}_0))}\B{x}_i
	\label{fullr}
	\end{equation}
	as an approximation of $\B{x}_T$, where $\beta$ is a parameter. Experiments in later sections were conducted either using the best response $\B{x}_0$ or the full response with $\beta = 100$ and performance is compared for several values of $n$ and $b$. \\
	\indent Another approach to get better approximations of $\B{x}_T$ is to solve a sequence of QUBOs with bias. More precisely, get an initial approximation of $\B{x}_T$ by following the previous procedure to produce $\B{x}_{T_1}$. Then modify the QUBO by adding a linear term $-\alpha \B{x}_{T_1}^t\B{x}$ where $\alpha > 0$ and find approximate solutions to 
	\begin{align}
		\underset{C_{n,b}}{\text{argmin }} \ \B{r}^t\B{x} + \B{x}^tQ\B{x} - \alpha \B{x}_{T_1}^t\B{x} = \underset{C_{n,b}}{\text{argmin }} \ (\B{r} - \alpha \B{x}_{T_1})^t\B{x} + \B{x}^tQ\B{x}
		\label{bias}
	\end{align} 
	By Cauchy-Schwartz, $\frac{\B{x}_{T_1}}{||\B{x}_{T_1}||} = \underset{||\B{x}||=1}{\text{argmin }} -\B{x}_{T_1}^t\B{x}$, thus in solving \ref{bias} the annealer is encouraged to produce solutions in the direction of $\B{x}_{T_1}$. The annealer produces a new lower-energy solution $\B{x}_{T_2}$ and this process can be repeated until the new solution no longer has lower energy than the previous. Experimental results in later sections contain data with $\alpha = 0$ and $.1$. Biasing is most helpful in the initial phase of the algorithm when it is iteratively producing solutions close to previous solutions. In later phases biasing is less useful, as will be evident in Figures 4 and 5.
	\subsection{An Iterative Descent Algorithm}
    Algorithms that solve continuous optimization problems rely on a good initial guess an an iterative descent rule. These tasks can be formulated as QUBO problems when trying to minimize the Rayleigh quotient over the unit sphere.
	\subsubsection{Obtaining an Initial Guess}
	Let $\lambda_1 \leq \lambda_2 \leq \ldots \leq \lambda_n$ be the eigenvalues of $A$. To get an initial approximation of $\lambda_1$, one can ask to solve
	\begin{align}
	\label{initapprox}
	\underset{\B{x} \in C_{n, b}}{\text{argmin }} \ \B{x}^tA\B{x}
	\end{align}
	as an approximation of 
	\begin{align}
	 	\underset{||\B{x}||=1}{\text{argmin }} \B{x}^tA\B{x}
	 	\label{bmatrix}
	\end{align}
	\indent An immediate problem is that if $A$ is positive definite, the solution to \ref{initapprox} is just $\B{x} = 0$. This can be remedied by replacing $A$ with $A - \lambda I_n$, where $\lambda \in (\lambda_i, \lambda_{i+1})$ for some $i$. The eigenvectors are unaffected, the eigenvalues can be recovered from the new matrix and the solutions to
	\begin{align}
	\label{iterativeEqn}
	\underset{\B{x} \in C_{n, b}}{\text{argmin }} \B{x}^t(A - \lambda I_n)\B{x}
	\end{align}
	tend to be long, nonzero vectors very close to the span of eigenvectors which have negative eigenvalues for $A - \lambda I_n$, namely $\B{v}_1, \ldots, \B{v}_i$. A good initial choice is the average of the eigenvalues $\lambda = \frac{tr(A)}{n}$ and as the algorithm progresses, $\lambda$ will decrease towards $\lambda_1$. To converge in fewer iterations, it's better to choose $\lambda$ close to, but greater than $\lambda_1$, as we will examine later. \\
	\indent These observations lead to the following iterative fixed-point method to produce a good initial guess for the lowest eigenvector. Initially solve \ref{iterativeEqn} with $\lambda = \frac{tr(A)}{n}$ to produce a guess $\B{v}_1$. Update $\lambda$ using the Rayleigh quotient $\lambda = \frac{\B{v}_1^t A \B{v}_1}{||\B{v}_1||^2}$ and solve \ref{iterativeEqn} again possibly using $\B{v}_1$ as a bias vector to produce a second guess $\B{v}_2$. Repeat until $\lambda$ is no longer decreasing.\\
	\indent For small matrices, say $10 \times 10$, this procedure often suffices to produce the lowest eigenvalue with 2-3 digits of accuracy and the corresponding eigenvector to within a distance of order $.1$ of the true eigenvector. The descent stage of the algorithm increases the precision to essentially arbitrary order.
	\subsubsection{Iterative Descent}
    Suppose we want to minimize a function $f: \R^m \rightarrow \R$, and let $\nabla f$ and $H(f)$ denote the gradient and Hessian of $f$, respectively.
	Starting with an initial guess $\B{x}_0$, a common strategy is to Taylor expand $f$ around $\B{x}_0$
	\begin{align}
	f(\B{x}) & = f(\B{x}_0) + \nabla f \cdot \delta + \delta^t \frac{H(f)}{2} \delta + o(||\delta||^3) \\
	\delta & := \B{x} - \B{x}_0
	\label{taylor}
	\end{align}
	and choose $\delta$ to minimize $\nabla f \cdot \delta$, which is gradient descent, or to minimize $\nabla f \cdot \delta + \delta^t \frac{H(f)}{2} \delta$, which includes second order methods such as Newton's method, BFGS, Newton Conjugate-Gradient, etc. Once a better solution $\B{x}_1 = \B{x}_0 + \delta$ has been found, Taylor expand around $\B{x}_1$ again and repeat. The proposed algorithm obtains a good descent direction by using a QUBO to find good approximate solutions to	
	$$\underset{\delta \in C_{n,b}}{\text{argmin }} \nabla f \cdot \delta + \delta^t \frac{H(f)}{2} \delta$$	
	Similar to Newton-CG and BFGS, this method requires computing, but not inverting, the Hessian matrix, and benefits from a line search which possibly increases the size of $\delta$. See \cite{Stoer} for more details on classical optimization algorithms and the benefits of line search. Here the line search step amounts to minimizing a quadratic, and so the optimal scaling can be directly computed. \\
	\indent If $k^{th}$ approximate solution $\B{x}_k$ is closer to the true solution than any point in the discretized cube, one cannot expect minimizing the QUBO to produce a better solution. A key part of the descent phase is enforcing a minimum step size in addition to the line search so that the  candidate $k+1^{st}$ solution is possibly \emph{worse} than that $k^{th}$. If the candidate solution is worse, the algorithm discards the candidate and replaces the discretized unit cube $C_{n, b}$ by a scaled-down discretized cube $t \cdot C_{n,b}$ where $t << 1$, which amounts to repeating the procedure outlined in section 2 with the precision vector $t \cdot \B{p}$. Once the discretized cube has been scaled down, the algorithm continues running until it needs to scale down the cube further, or exits having achieved the desired accuracy.

\subsection{The Algorithm}
\label{sec:alg}
With the key ingredients covered, we are in a position to present the algorithm. As a reminder, at each step the objective function is of the form $f(\B{x}) = \B{x}^t(A-\lambda)\B{x}$, whose gradient and Hessian can be calculated as $\nabla f = 2(A-\lambda)\B{x}^t$, $H(f) = 2(A - \lambda)$, and these formulas are implicitly used in the descent phase of the algorithm. At several stages, the algorithm solves optimization problems of the form $\underset{ \B{x} \in t \cdot C_{n,b}}{\text{argmin}} \ \B{r}^t\B{x} + \B{x}^tA\B{x}$. These are turned in to QUBOs as explained in section \ref{sec:background}, and annealers are used to minimize the QUBOs that appear, possibly using full responses or biasing. In subsequent section the effects of biasing, full responses and other parameters will be analyzed.\\
\begin{algorithm} [H]
\caption{Controlled Precision QUBO-based Algorithm to Compute Eigenvectors of Symmetric Matrices}
\label{alg}
\begin{algorithmic}

\STATE Inputs: Symmetric $n \times n$ matrix $A$,  bits for precision vector $b$, desired precision $\e_{tol}$
\STATE Outputs: Approx smallest evec, eval $\B{v}, \lambda$ within $\e_{tol}$ of true values
\STATE $\lambda \gets \frac{tr(A)}{n}$
\STATE$H \gets A - \lambda I_n$ \COMMENT{Enforcing Indefiniteness}
\STATE$\B{v} \gets \underset{C_{n,b}}{\text{argmin}} \ \B{x}^t H \B{x}$ \COMMENT{Initial Guess Phase} 
\STATE$\B{v} \gets \frac{\B{v}}{||\B{v}||}$ 
\WHILE{$\B{v}^t A \B{v} < \lambda$ }
 	\STATE$\lambda \gets \frac{\B{v}^tA\B{v}}{||\B{v}||^2}$ 
 	\STATE$H \gets A - \lambda\cdot I_n$
 	\STATE$\B{v} \gets \underset{C_{n, b}}{\text{argmin }} \B{x}^tH\B{x}$
 	\STATE$\B{v} \gets \frac{\B{v}}{||\B{v}||}$ 
\ENDWHILE  
\STATE$precision \gets .1$ \COMMENT{Descent Phase}
\WHILE{$precision > \e_{tol}$}
    \STATE$H \gets A - \lambda I_n$
    \STATE$\delta \gets \underset{precision \cdot C_{n, b}}{\text{argmin }} 2\B{v}^tH\delta +  \delta H \delta$ \COMMENT{Getting Descent Direction} 
    \STATE$\delta \gets \delta - \langle \B{v}, \delta \rangle \delta$ \COMMENT{Orthogonalizing $\delta, \B{v}$}
    \STATE$t_{min} = \max\{-\B{v}^tH \delta / (\delta^t H \delta),1\}$ \COMMENT{Line search step}
    \STATE$\delta \gets t_{min} \cdot \delta$
    \IF{$\frac{(\B{v} + \delta)^t A (\B{v} + \delta)}{||\B{v} + \delta||^2} < \lambda$} \COMMENT{Checking if solution improves}
        \STATE$\B{v} \gets \frac{\B{v} + \delta}{||\B{v}+\delta||}$
        \STATE$\lambda \gets \B{v}^t A \B{v}$
    \ELSE
        \STATE$precision \gets .1 \cdot precision$ \COMMENT{Increasing precision otherwise}
    \ENDIF  
\ENDWHILE
\RETURN \ $\B{v}, \lambda$
\end{algorithmic}
\end{algorithm}

\indent Two steps merit a bit more explanation. Replacing $\delta$ by $\delta - \langle \B{v}, \delta \rangle \delta$ forces $\B{v}, \delta$ to be orthogonal. Since the Rayleigh quotient needs to be optimized over the sphere, the update direction $\delta$ should be tangent to the sphere at $\B{v}$, and the tangent space of the sphere at $\B{v}$ is precisely the set of vectors orthogonal to $\B{v}$. Second, the scaling $\delta$ is computed as $t_{min} = \max\{-\B{v}^tH \delta / (\delta^t H \delta),1\}$. The expression $-\B{v}^tH \delta / (\delta^t H \delta)$ is the line search step coming from minimizing the quadratic $(\B{v} + t_{min}\delta)^t H(\B{v} + t_{min}\delta)$. Strictly speaking this quadratic only has a minimum when $\delta^t H \delta$ is positive, and in practice when using this algorithm it almost always is, and if not, set $t_{min} = 1$. A minimum scaling $t_{min} \geq 1$ is enforced so that the candidate update $\B{v} + t_{min}\delta$ possibly overshoots the exact solution, resulting in a worse estimate of the lowest eigenvector. Overshooting is an indication that the discretized cube is no longer fine enough to produce better solutions, and so the candidate update is discarded and the discretized cube is scaled down. Intuitively the scaling at each step should be about the order of $\frac{1}{2^{b-1}}$, and the numerical experiments below all use $.1$. \\
	\indent Oftentimes in practice one wishes to solve a generalized eigenvalue problem of the form $A \B{v} = \lambda B\B{v}$. In the case when $A, B$ are symmetric and $B$ is strictly positive definite, the smallest generalized eigenvalue minimizes the generalized Rayleigh quotient
	\begin{align}
		\lambda = \min_{||\B{x}||=1} \frac{\B{x}^tA\B{x}}{\B{x}^tB\B{x}} \ \ \B{v} = \underset{||\B{x}||=1}{\text{argmin }} \frac{\B{x}^tA\B{x}}{\B{x}^tB\B{x}}
	\end{align}
	The following small changes solves the generalized eigenvalue problem, again to essentially arbitrary precision. First, instead of initializing $\lambda$ as $\frac{tr(A)}{n}$, one can generate a random unit vector $\B{w}$ (or use a specified vector) and initialize $\lambda = \frac{\B{w}^tA\B{w}}{\B{w}^tB\B{w}}$. Second, replace every Rayleigh quotient with the corresponding generalized Rayleigh quotient. Lastly, instead of updating $H$ as $H = A - \lambda I_n$, update as $H = A - \lambda B$, as the latter preserves the $B$-eigenspectrum of $A$, while the former does not. \\
	\indent We conclude by emphasizing that this algorithm reaches arbitrary precision without increasing the size of the QUBOs, all of which involve $n \cdot b$ binary variables. Additionally, all quadratic problems are of the form $\B{r}^t\B{x} + \B{x}^t(A - \lambda)\B{x}$ where $A$ is fixed, implying that the potential non-zero coefficients of the QUBO do not change (examine formula \ref{finalQUBO}).

\section{Experimental results}
\label{sec:results}
The algorithm and its variants are tested on a class of random matrices of varying sizes. 
For each experiment, the algorithm ran until the Rayleigh quotient of the approximate eigenvector was within $10^{-8}$ of the true value.
 \subsection{Basic Performance}
\begin{figure}[H]
 \label{fig:basicperformance}
 	\centering
 	\subfigure{\includegraphics[width=6cm]{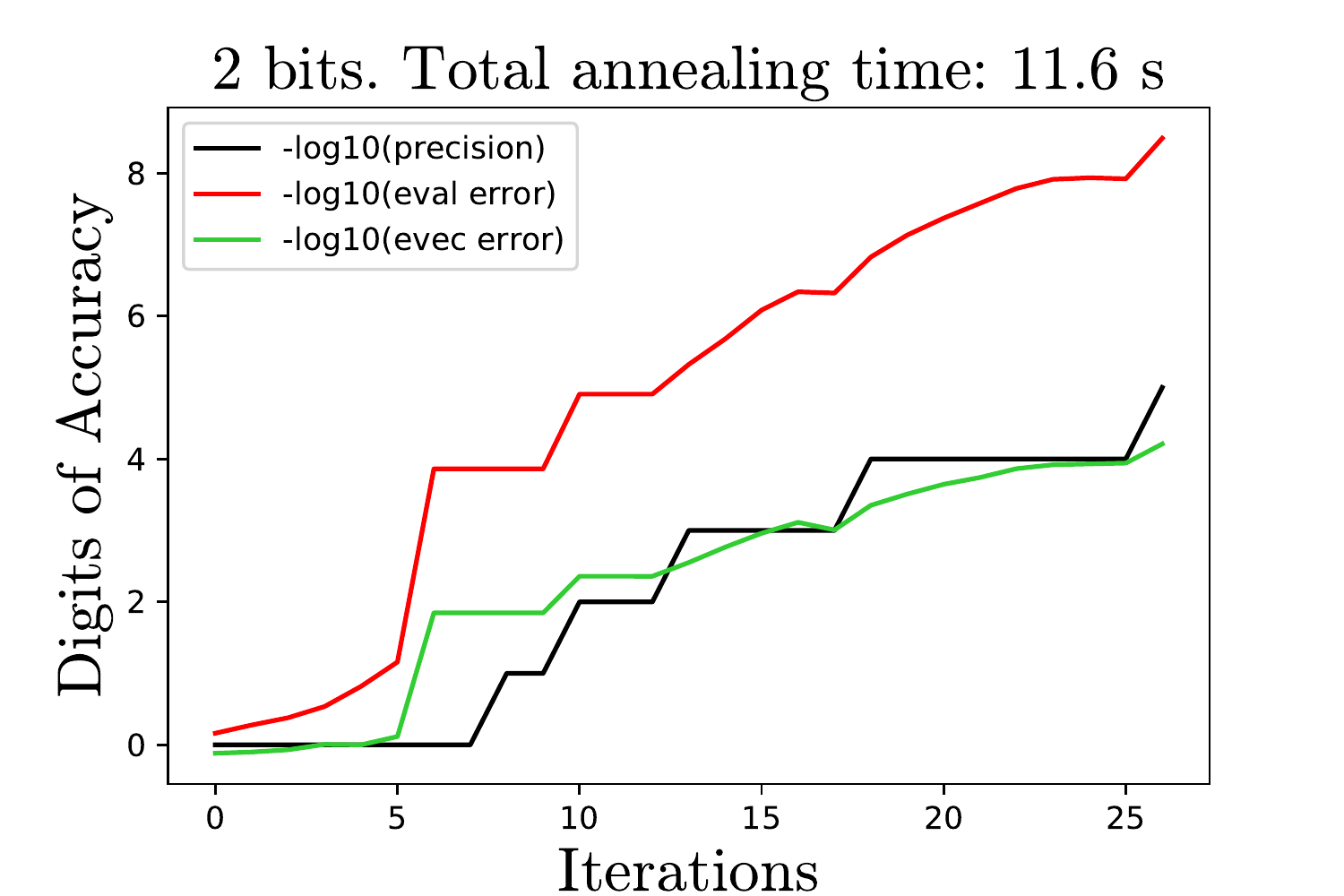}} 
 	\subfigure{\includegraphics[width=6cm]{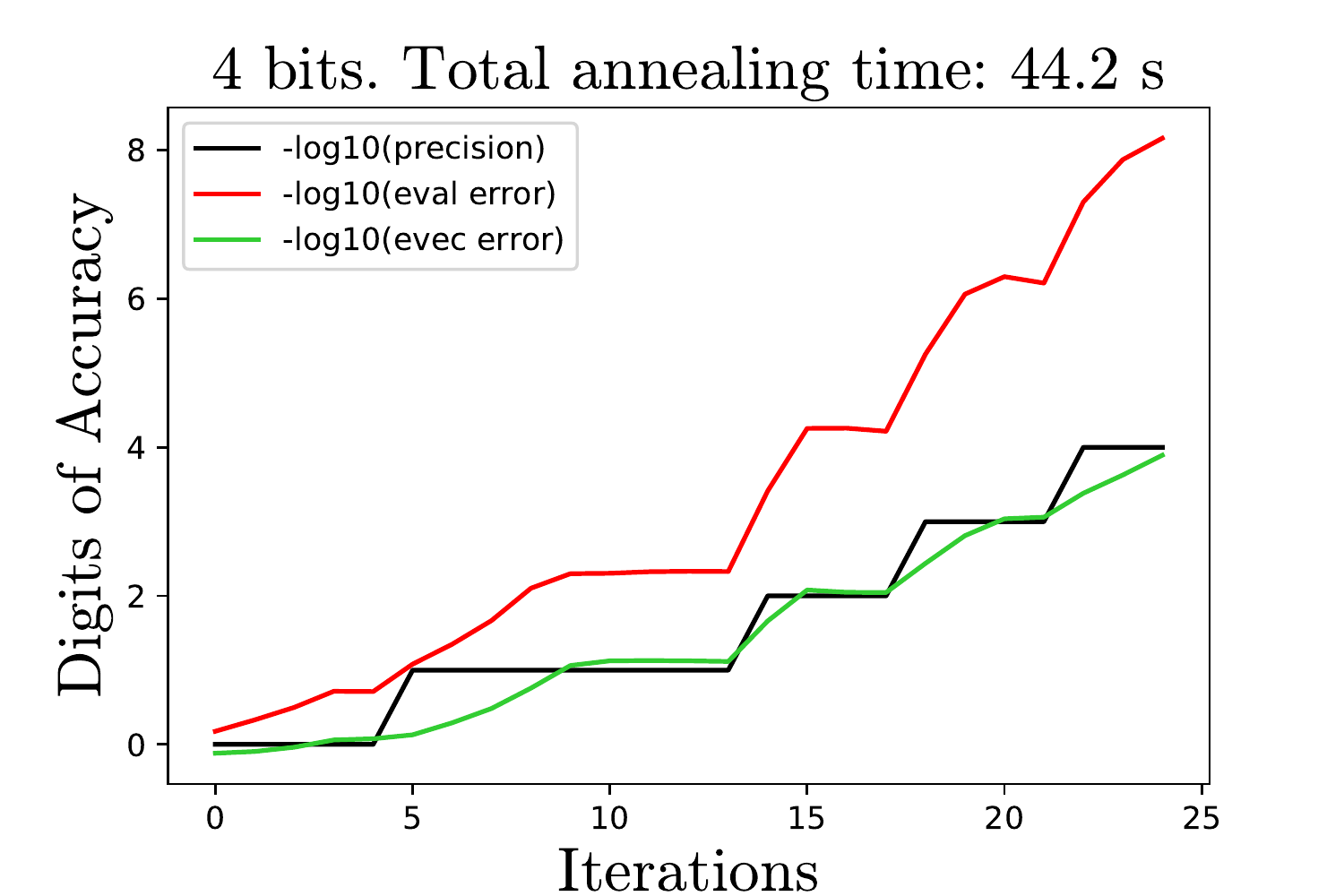}} \\
 	\subfigure{\includegraphics[width=6cm]{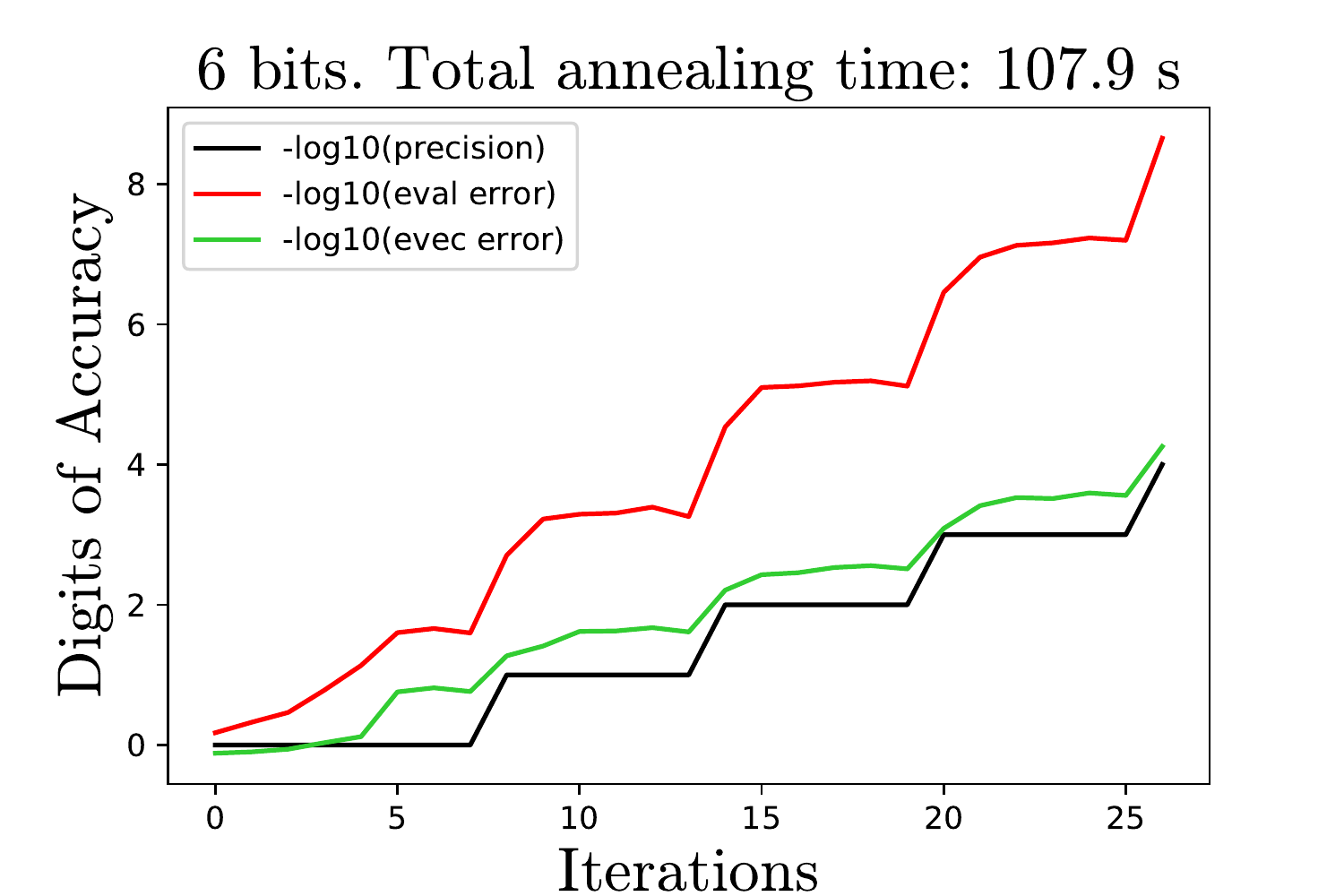}} 
 	\subfigure{\includegraphics[width=6cm]{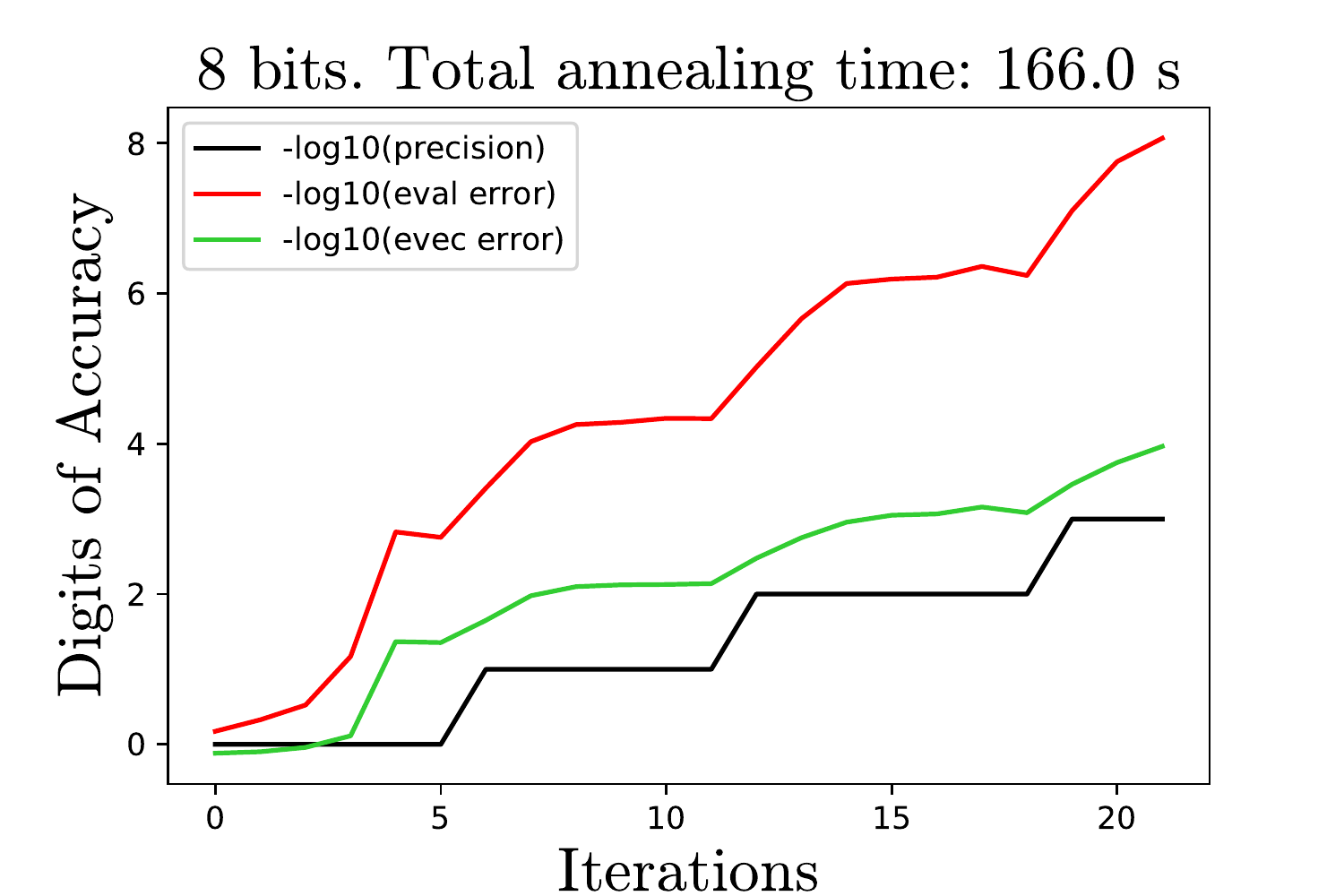}}
 	\caption{Digits of accuracy plotted against number of iterations. \emph{Total annealing time} is the total amount of time the algorithm spends on performing SA, since SA dominates the cost of the method. \emph{Evec error} is the distance of the computed vector from the true unit eigenvector. \emph{Precision} refers to the scaling applied to the discretized cube at each iteration. The initial guess phase corresponds to a precision of 1. Observe that whenever the error increases, the algorithm responds by increasing the precision, which often gives large accuracy gains within the subsequent 2-3 iterations.}
 \end{figure}

\begin{figure}[H]
    \label{fig:spaceshuttle}
	\centering 
	\subfigure{\includegraphics[width=6cm]{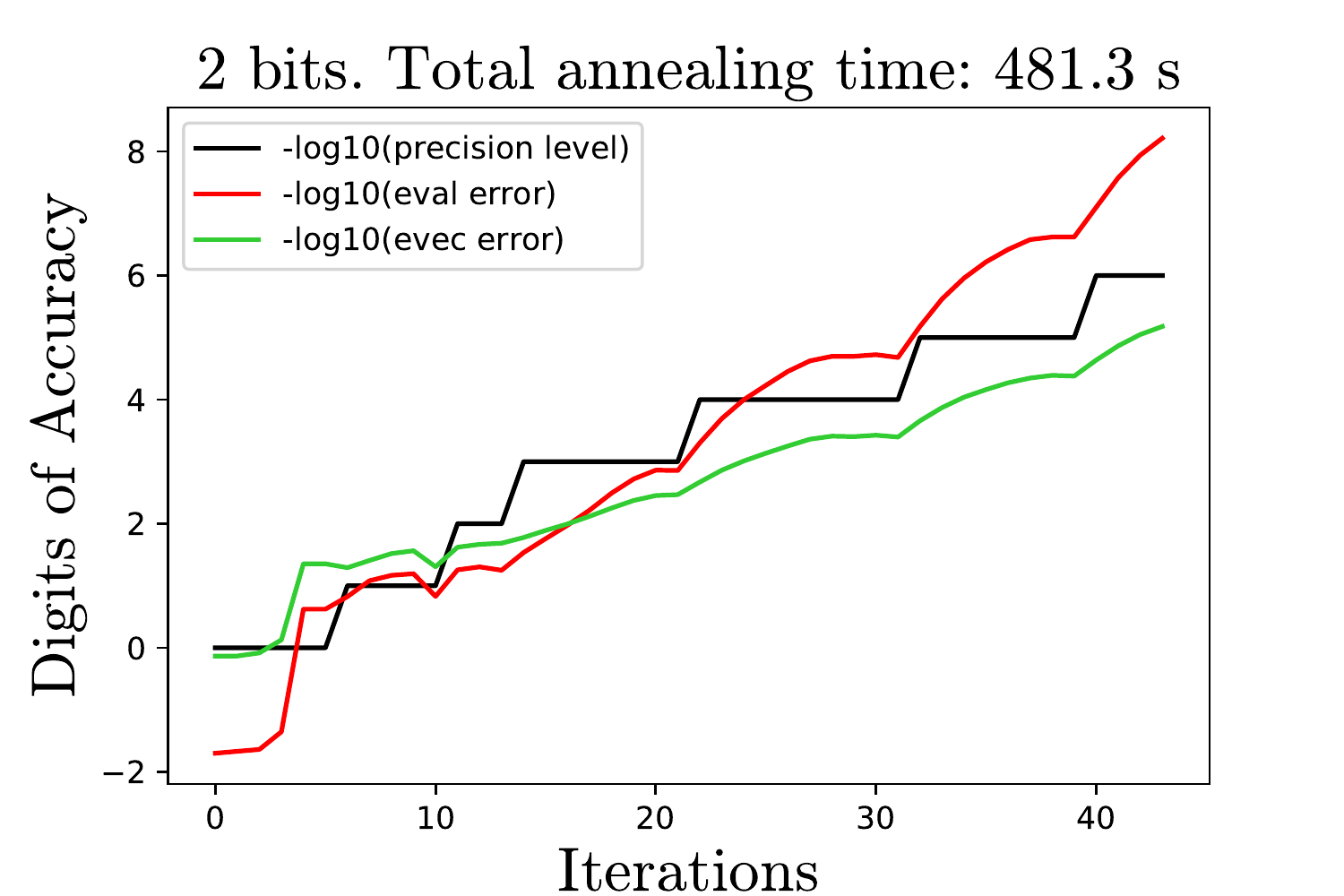}} 
	\subfigure{\includegraphics[width=6cm]{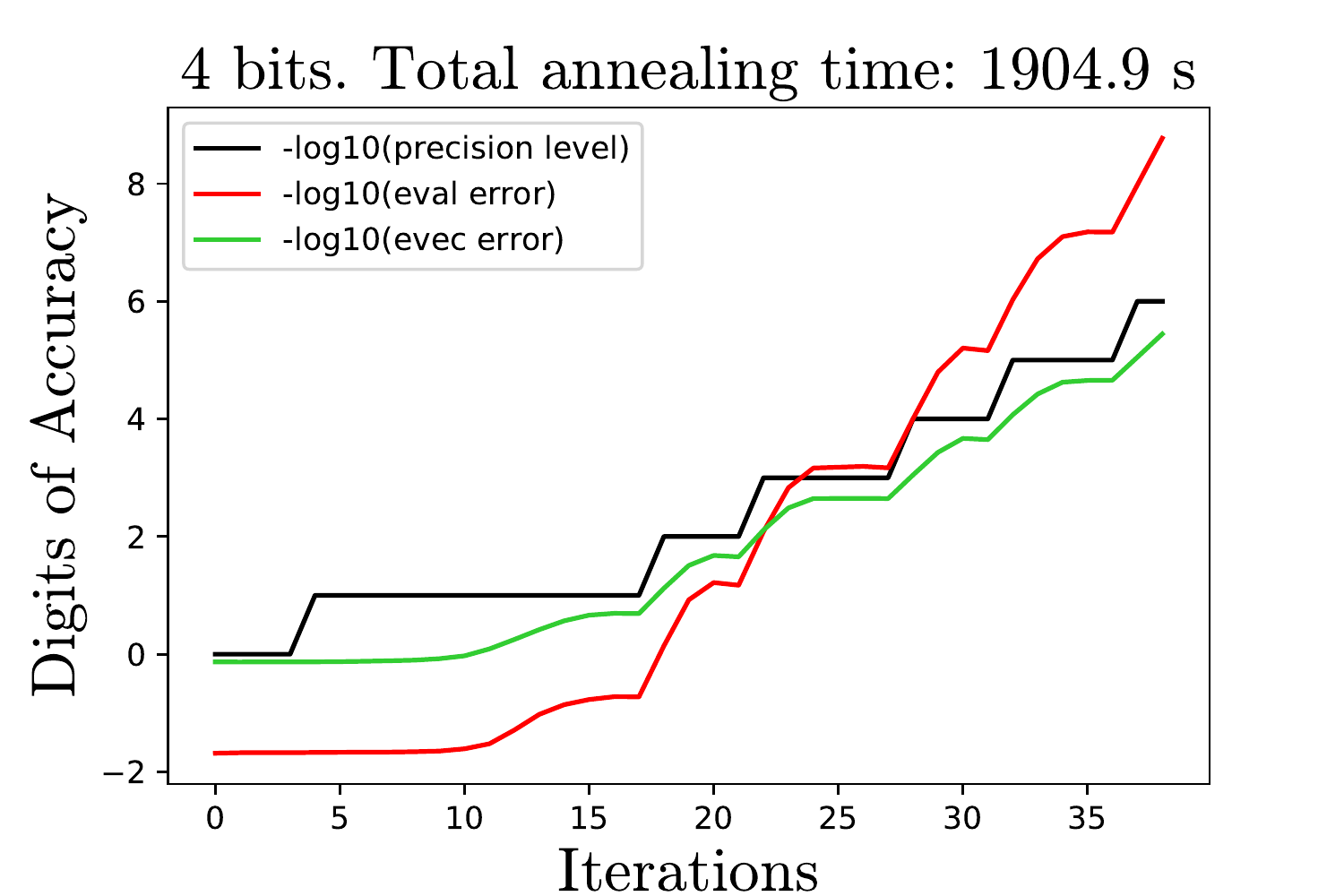}} 
	\caption{Error plot for $560 \times 560$ Space Shuttle Control Matrix}
\end{figure}

First, we demonstrate the convergence as a function of the number of iterations using example matrices from the TAMU SuiteSparse collection \cite{UFL}. Figure \ref{fig:basicperformance} shows performance on the breasttissue\_10NN matrix, a weighted graph adjacency matrix of size $106\times106$ for $2, 4, 6$ and $8$ bits using best response and no biasing.
 Interestingly using fewer bits gives less time to reach desired accuracy despite requiring more iterations. A $log-log$ regression on the MP matrices (see next section) gives that the anneal time grows like $(n\cdot b)^{1.57}$ and the number of iterations grows like $n^{.44}b^{-.32}$ so the total time is roughly $n^2b^{1.2}$. The algorithm works on even larger matrices, as is demonstrated in Figure \ref{fig:spaceshuttle} using the spaceShuttleEntry\_1 matrix, a $560 \times 560$ control matrix.

\begin{figure}[H]
    \label{fig:genEval}
	\centering 
	\subfigure{\includegraphics[width=6cm]{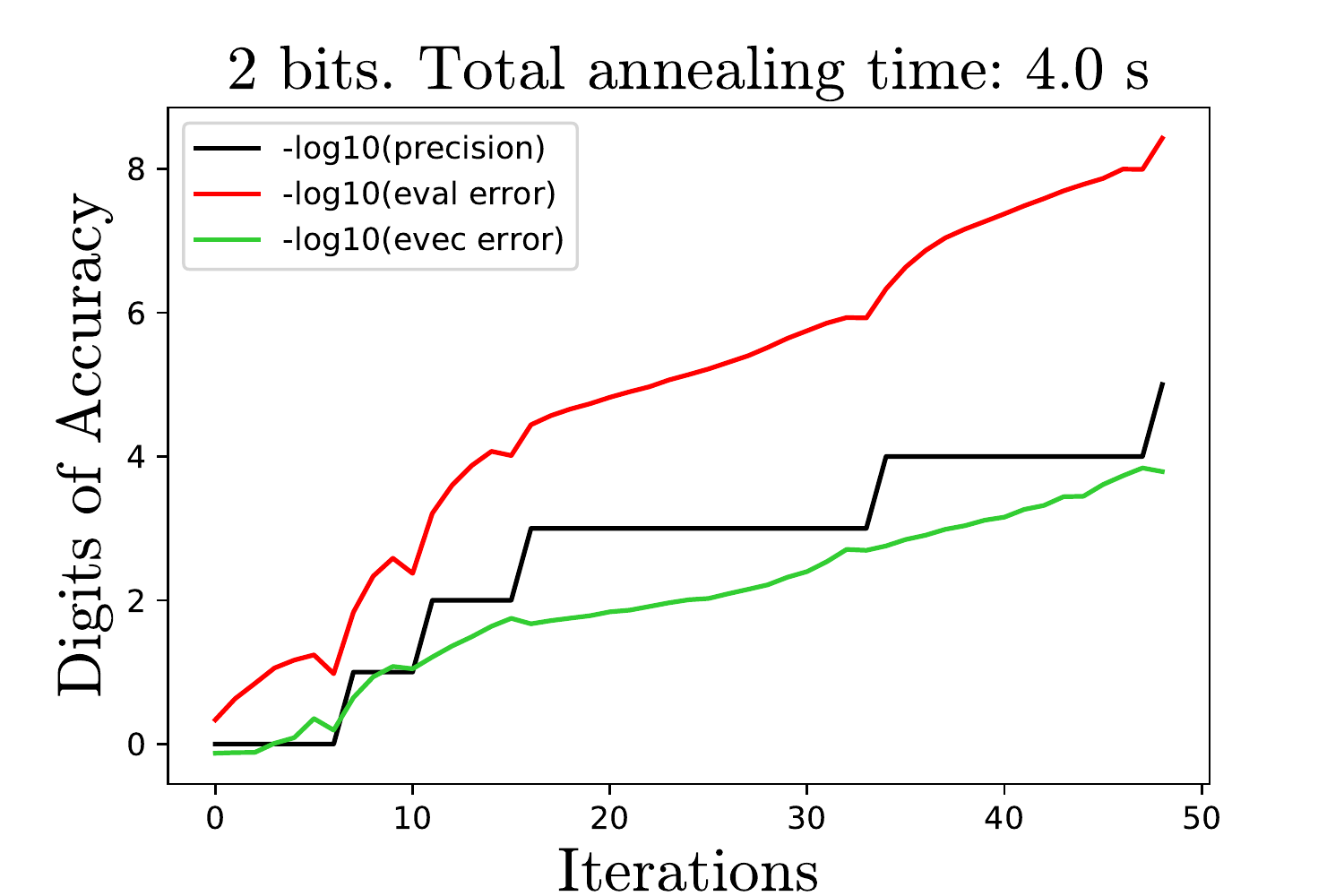}} 
	\subfigure{\includegraphics[width=6cm]{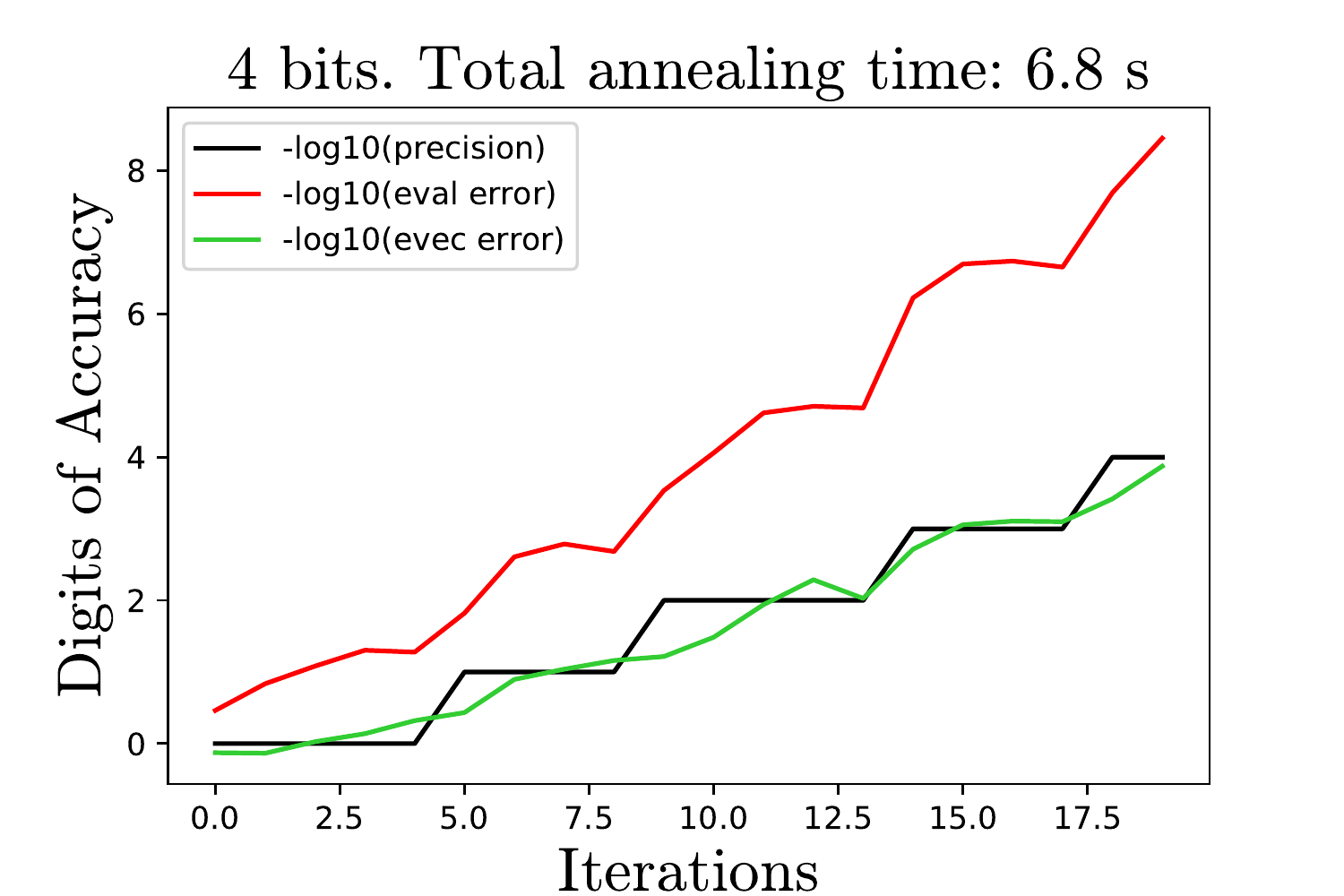}} \\
	\subfigure{\includegraphics[width=6cm]{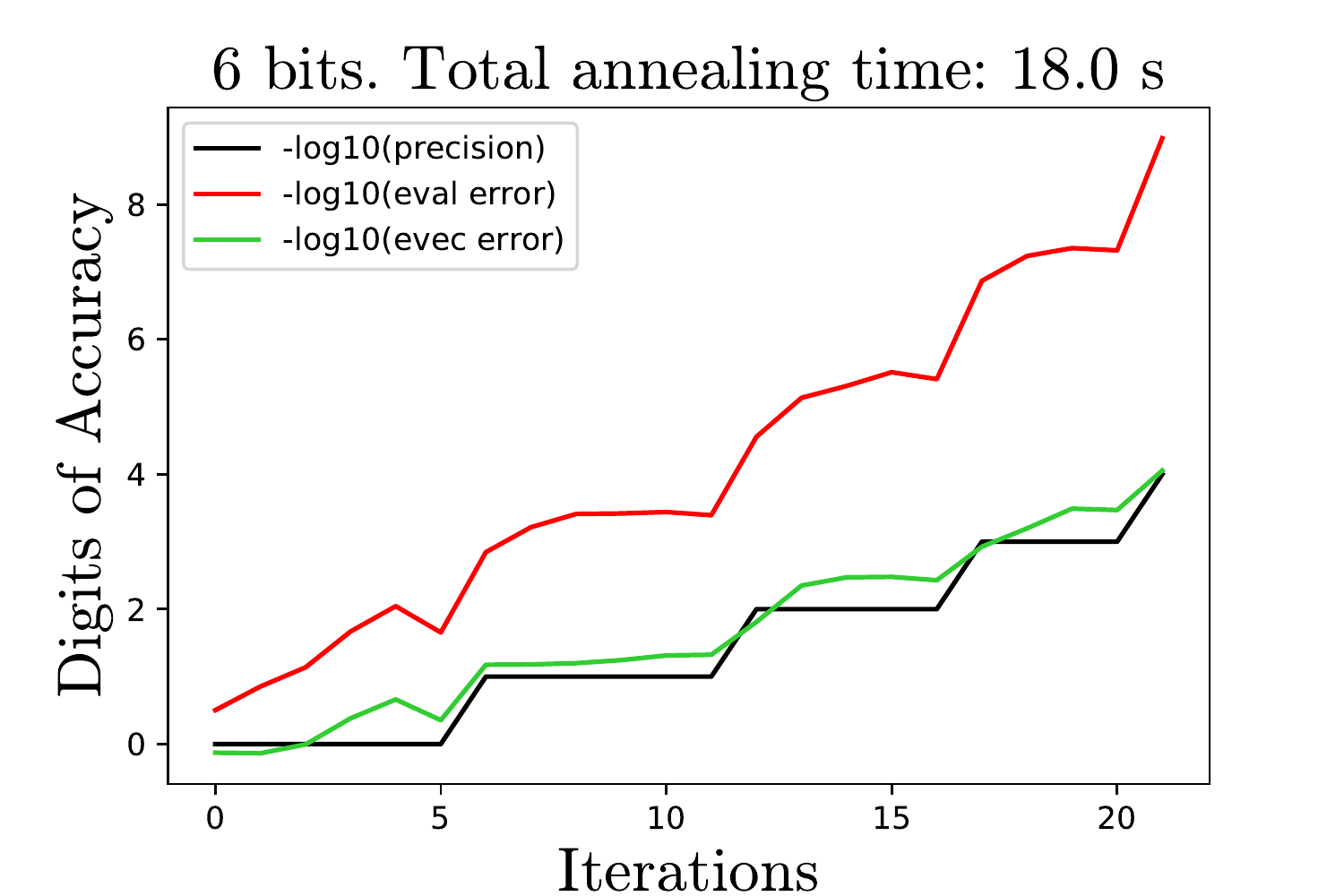}} 
	\subfigure{\includegraphics[width=6cm]{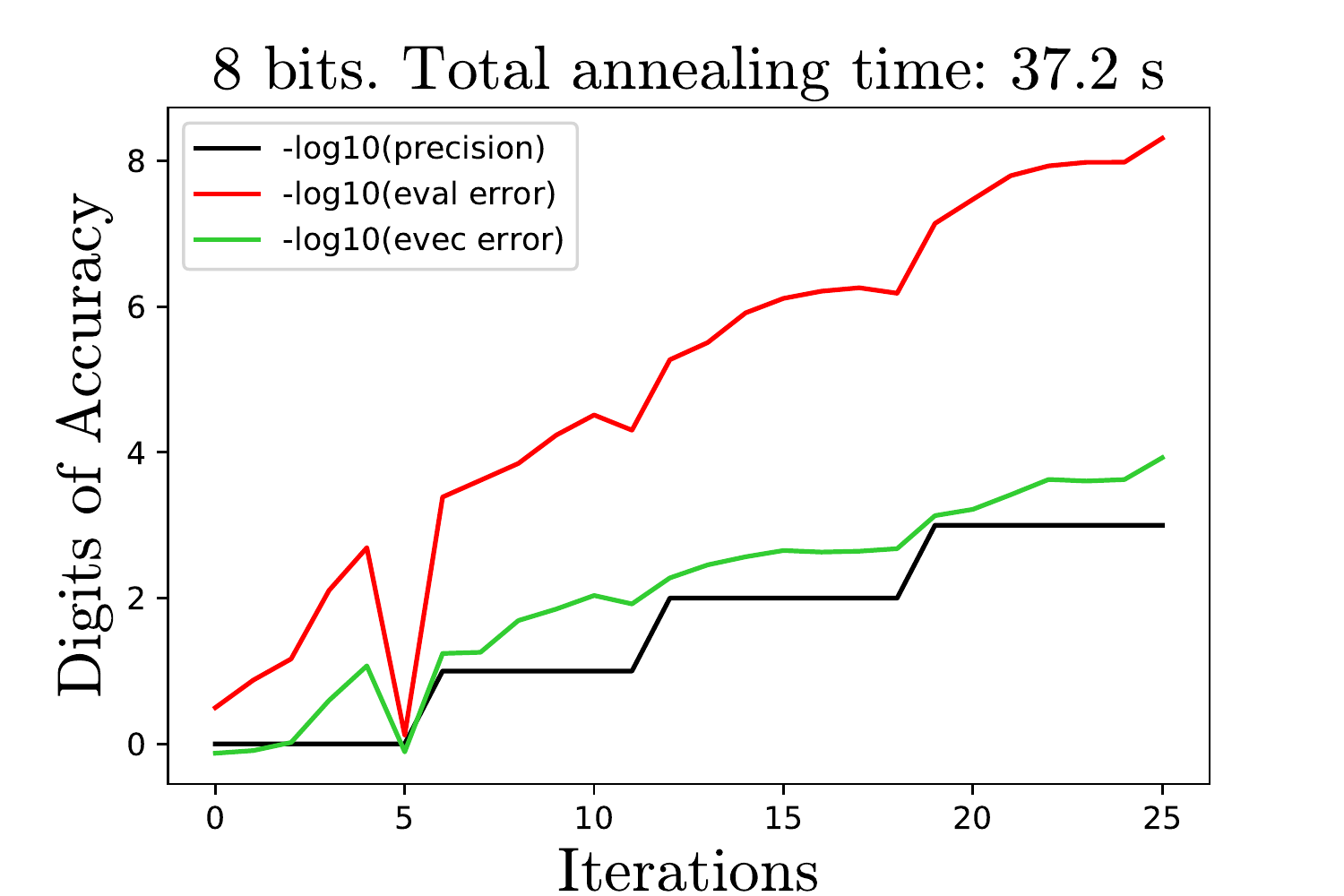}}
	\caption{Error plots for Generalized Eigenvalue Problem on two $48\times48$ mesh matrices}
\end{figure}

\indent Figure \ref{fig:genEval} demonstrates the performance for the generalized eigenvalue problems using mesh1em1 as the $A$ matrix and meshe1 as the $B$ matrix, two $48 \times 48$ matrices from the SuiteSparse database.

\indent In order to try to get algorithms that run as fast as possible, one might ask if it is possible get the algorithm to work using 1 bit of precision. With the current scheme this cannot be done. However, by reformulting the problem as an Ising instead of QUBO, one can indeed use only one bit precision. Ising problems are of the form
$$\underset{\B{x} \in \{-1, 1\}^n} {\mathrm{argmin}} \B{h}^t\B{x} + \B{x}^t Q \B{x}$$
the main distinction from QUBOs being the spin variables $\pm 1$. QUBOs or Ising problems are mathematically equivalent, and most annealers are capable of solving either.\\
\indent Using the Ising formulation, its possible to mimic the same algorithm, which works well on very small matrices. However, for larger matrices, such as for the $106 \times 106$ weighted adjacency matrix the single-bit version of the algorithm takes longer than using two bits, taking $32.3$ seconds and requiring over 400 iterations (compare with Figure \ref{fig:basicperformance}). An educated guess for why this might happen follows. Since the solutions produced by Ising problems have coordinates that are all non-zero and of the same magnitude, if the algorithm has already produced a solution whose $k^{th}$ coordinate is close to the true value, the added solution from the Ising problem will force that coordinate away from the optimal value. Using two bits is effective because the solutions can have coordinates that are positive, negative or zero. 

\subsection{Analysis of Parameters}
\begin{figure}[H]
 \label{fig:guessPhaseError}
	\centering 
	\subfigure{\includegraphics[width=6cm]{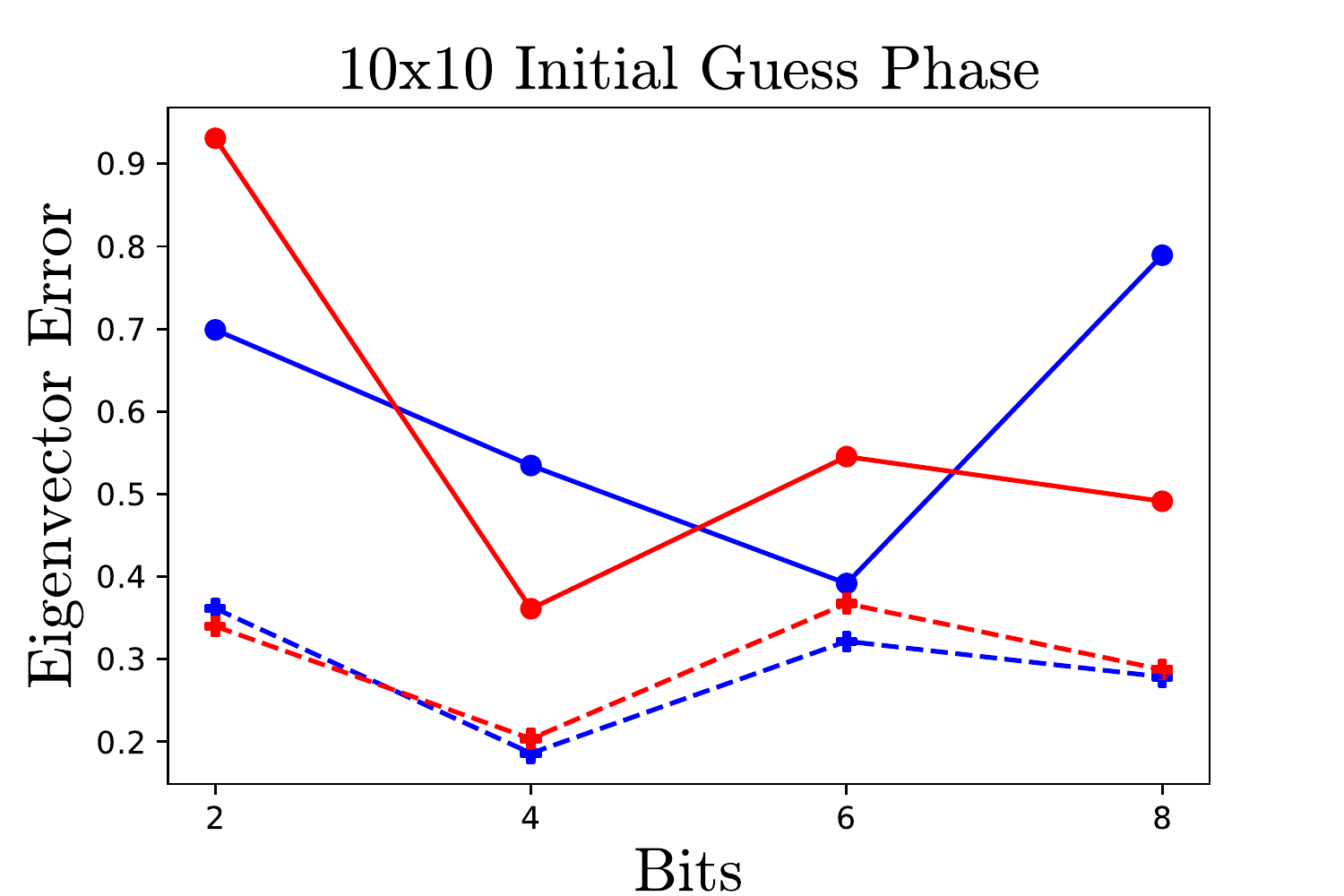}} 
	\subfigure{\includegraphics[width=6cm]{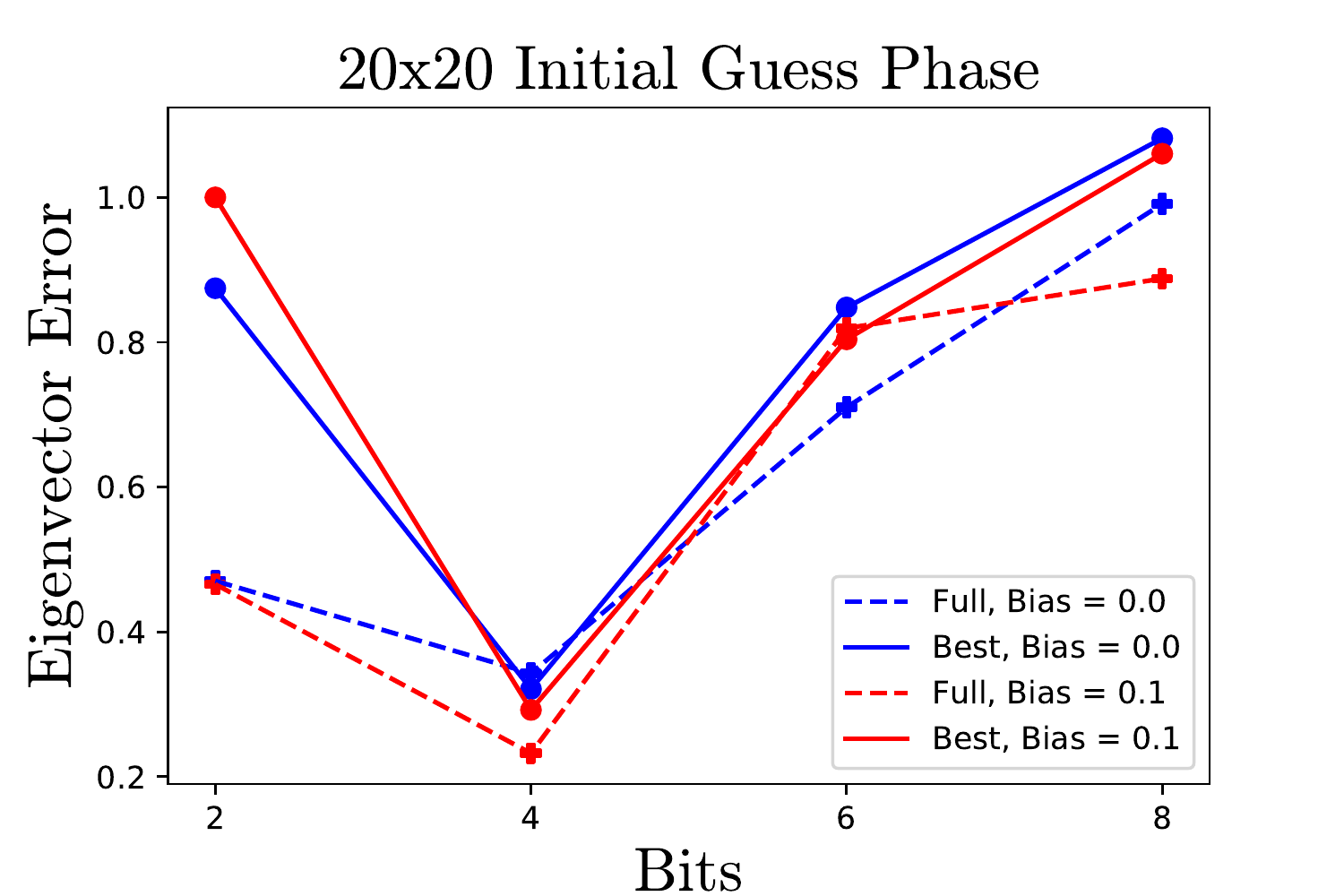}} 
	\caption{Eigenvector Error for MP matrices at the end of initial guess phase. Observe that for small QUBOs the full response decreases the error significantly.}
\end{figure}

\begin{figure}[H]
\label{fig:diffParams}
	\centering
	\subfigure{\includegraphics[width=6cm]{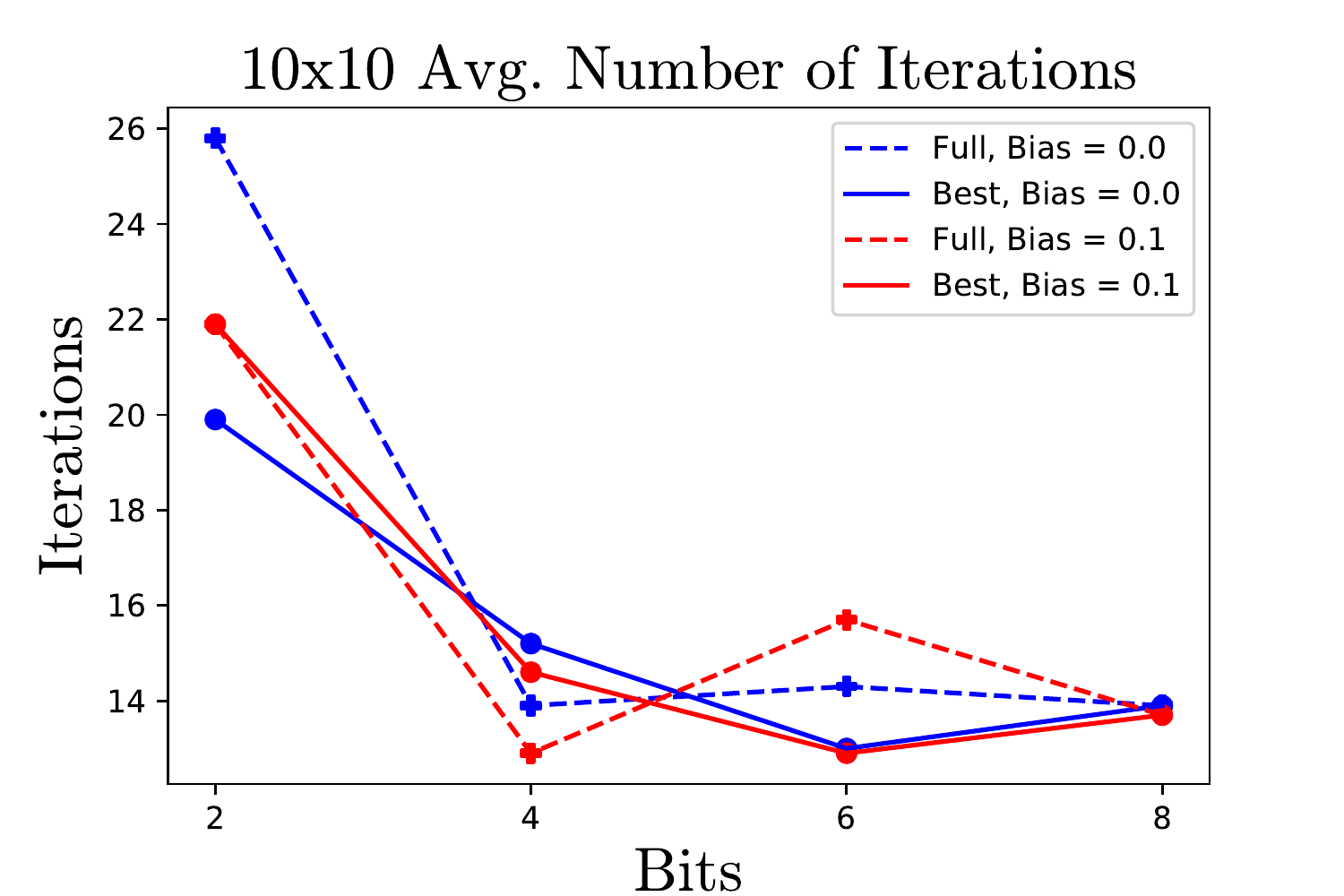}} 
	\subfigure{\includegraphics[width=6cm]{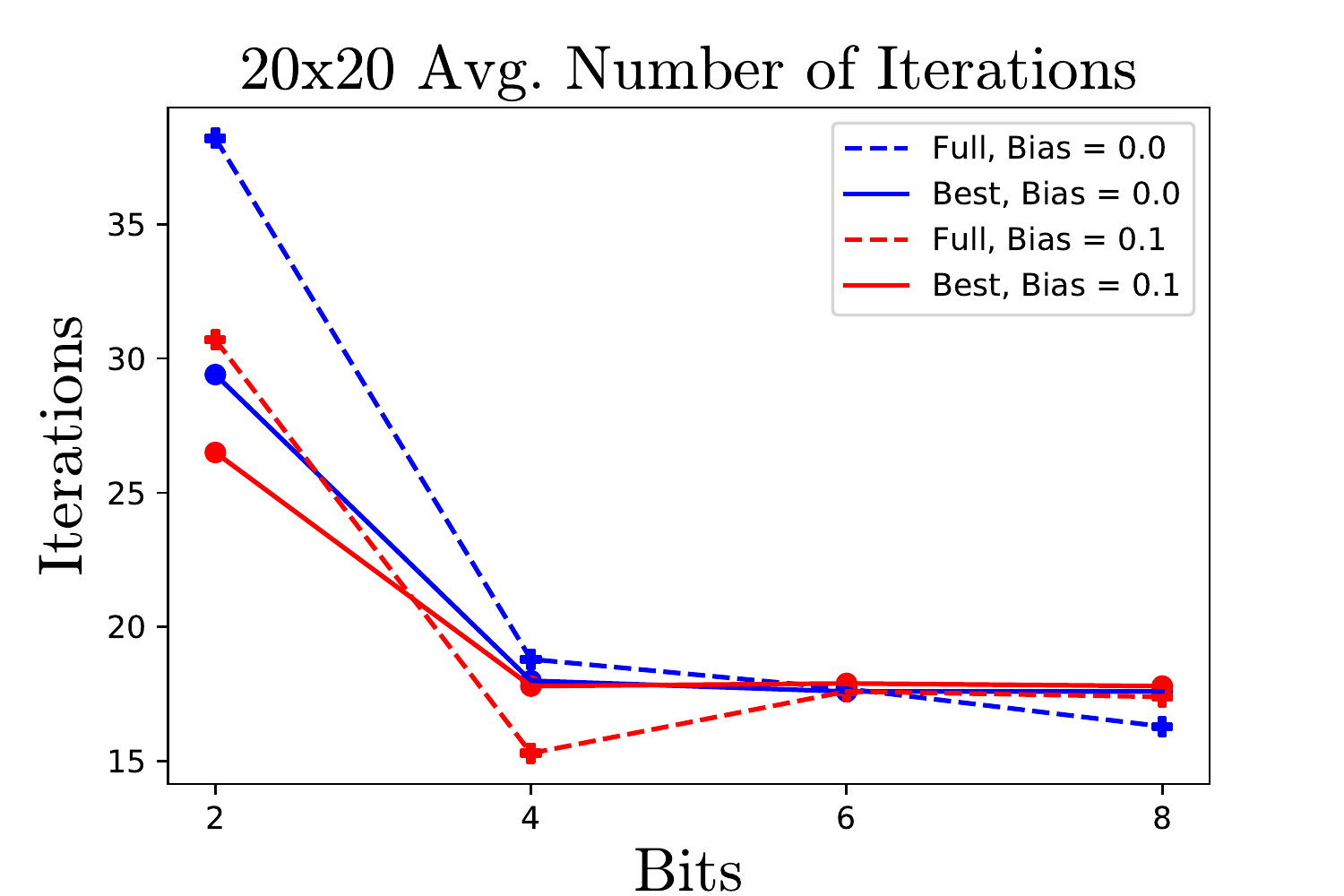}} 
	\caption{Average number of iterations required for MP matrices for different parameters.}
\end{figure}

To demonstrate the effect of biasing and full response parameters, the algorithm is tested on small matrices of sizes $3$, $10$ and $20$ with number of bits $b = 2, 4, 6$ and $8$. In the interest of not overwhelming the reader with plots and data, only the data for matrices of size $10$ and $20$ is displayed. We analyze the error at the end of the initial guess phase, and the average number of iterations each method requires. For each choice of size, bits and parameters, 10 Marchenko-Pasture matrices \cite{Marchenko_1967} with parameter $\lambda = .3$ are generated and the average errors at the end of the descent phase is recorded. Ideally this initial phase should end with the smallest possible error before beginning the descent phase. Towards this end, taking full responses (equation \ref{fullr}) and biases (equation \ref{bias}) can be very beneficial. However, this benefit fades as the sizes of the QUBOs increase as one can see from Figures \ref{fig:guessPhaseError} and \ref{fig:diffParams}.

\begin{figure}[H]
	\centering
	\subfigure{\includegraphics[width=6cm]{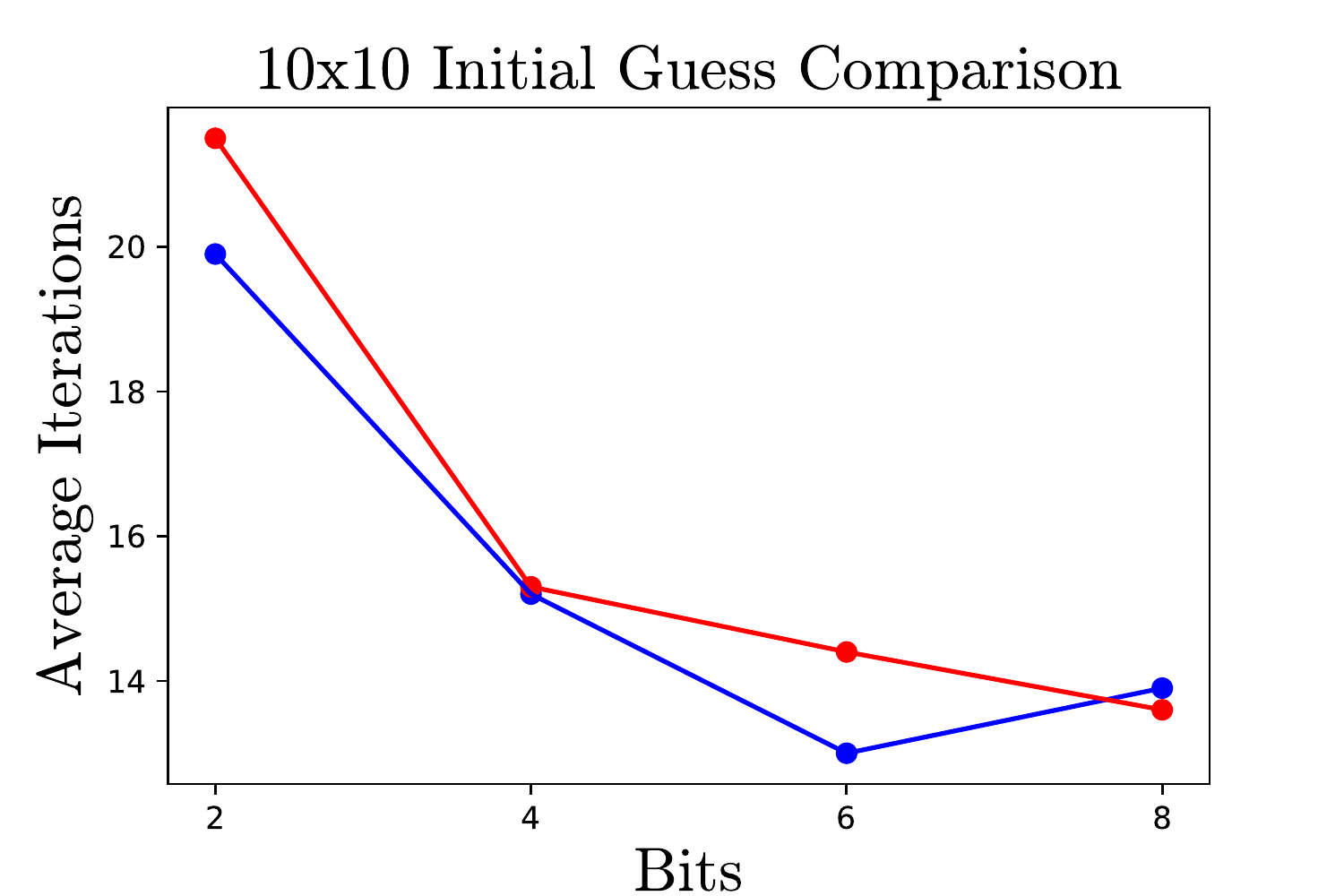}} 
	\subfigure{\includegraphics[width=6cm]{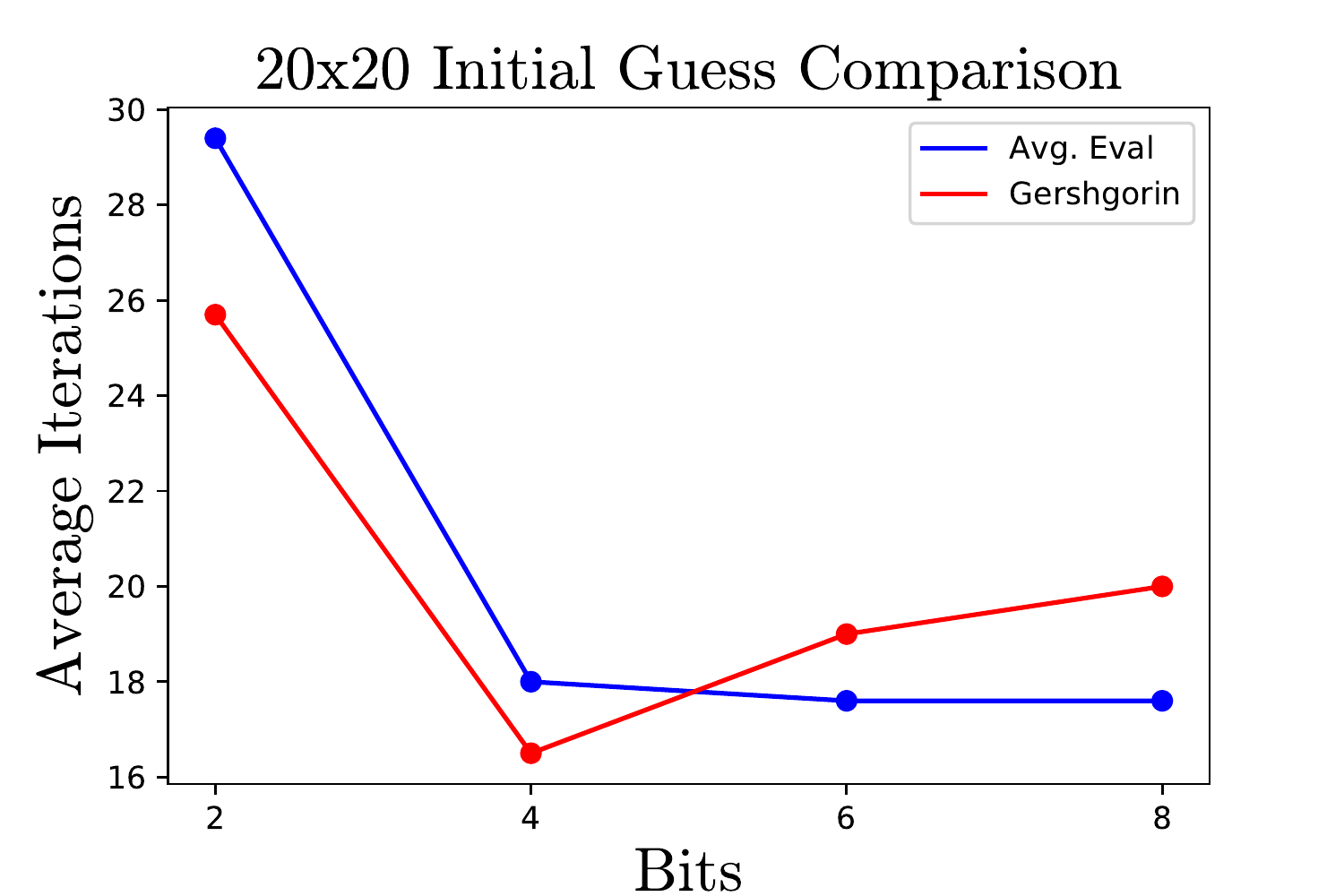}} 
	\caption{Total Iterations required for different initializations of $\lambda$}
	\label{init}
\end{figure}

\indent The choice to initialize $\lambda$ as $\frac{tr(A)}{n}$ is motivated by a desire to produce an initial guess which is close to, but greater than, the true lowest eigenvector.  To demonstrate this effect on $10 \times 10$ and $20 \times 20$ matrices, we compare performance initializing as $\frac{tr(A)}{n}$, which is the average of all the eigenvalues, against initializing as the highest Gershgorin bound, which upper bounds the maximum eigenvalue \cite{Gersh}. Choosing an initialization closer to the true eigenvalue often leads to fewer iterations, although the difference is somewhat small and fades as the number of bits increases, as seen in Figure \ref{init}.

\subsection{Gap Size Analysis}
\begin{figure}[H]
\label{fig:gapsize}
	\centering
	\subfigure{\includegraphics[width=6cm]{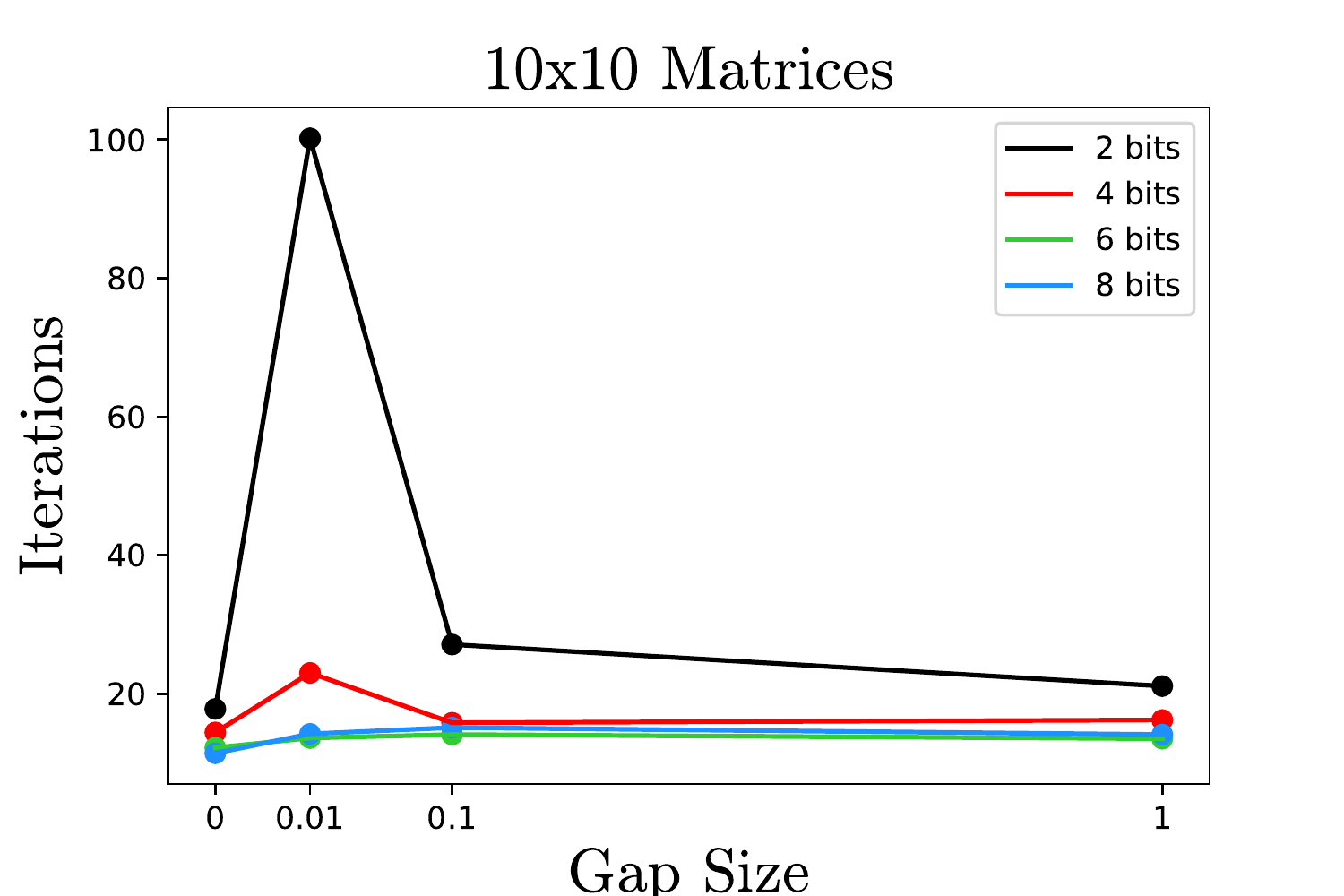}} 
	\subfigure{\includegraphics[width=6cm]{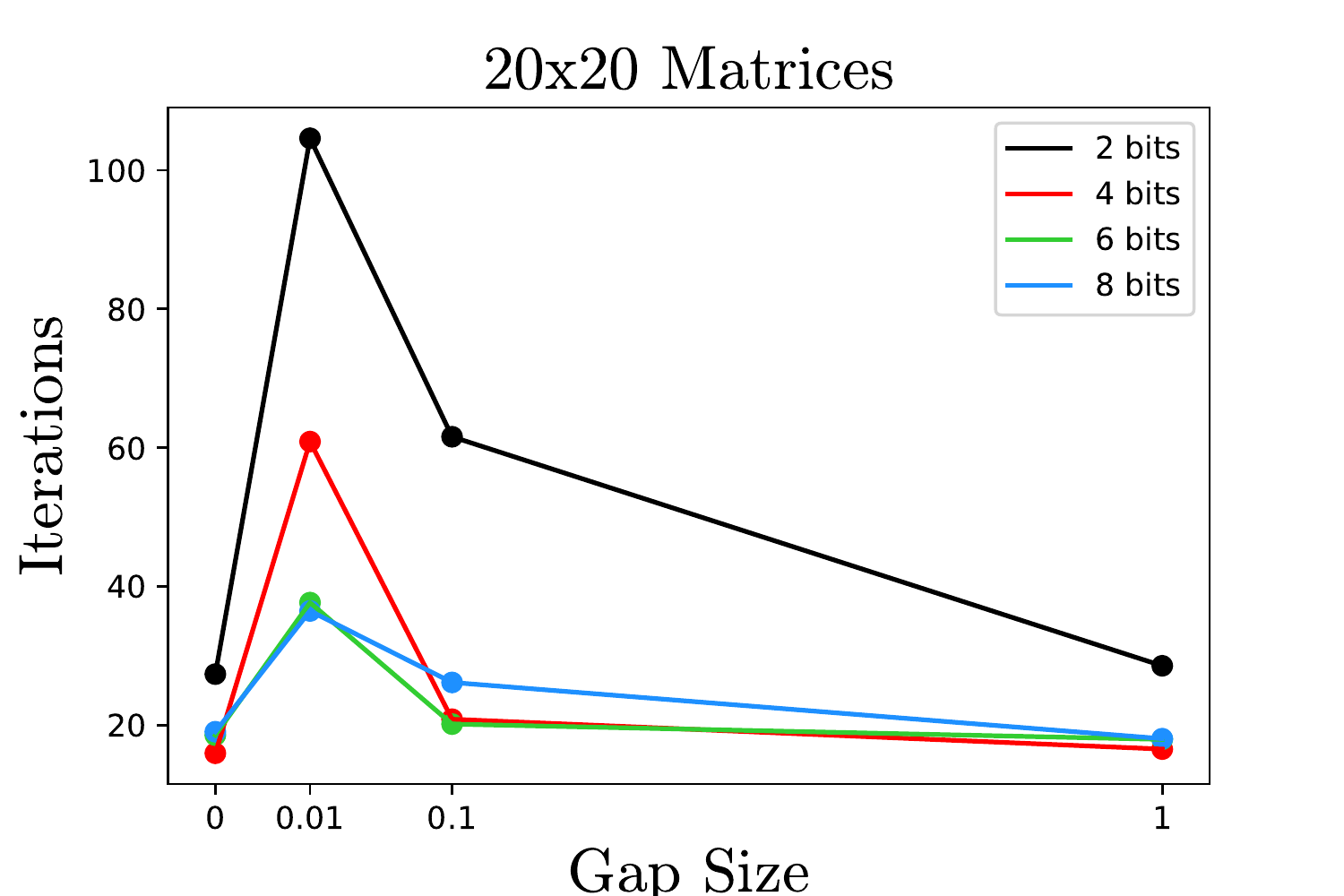}}  
	\caption{Average number of iterations on 30 samples as a function of the gap size $|\lambda_1 - \lambda_2|$. The smaller the gap, the more iterations required, especially when the number of bits is small. In the extreme case where the gap is $0$ and the lowest eigenvalue appears with multiplicity, the algorithm is actually faster in the sense that fewer iterations are needed for the computed eigenvalue to approximate the true eigenvalue.}
\end{figure}

Here we analyze the effect of the spacing between eigenvalues. In particular, the gap $|\lambda_1 - \lambda_2|$ can significantly affect the number of iterations required to reach a given precision. For this experiment, given a gap size $g = |\lambda_1 - \lambda_2|$, an orthogonal matrix $U$ is chosen at random with respect to the Haar measure using the SciPy implementation of \cite{HaarMeasure}. The algorithm is then analyzed on the matrix $U^t \text{Diag}(0, g, 1, \ldots, n-2) U$. As Figure \ref{fig:gapsize} demonstrates, as the gap size decreases the algorithm takes longer to achieve a given accuracy. The exception is when the gap is 0, and the smallest eigenvalue appears with multiplicity. In this case the algorithm actually requires fewer iterations.

\begin{figure}[H]
\label{fig:gapErrorPlots}
	\centering
	\subfigure{\includegraphics[width=6cm]{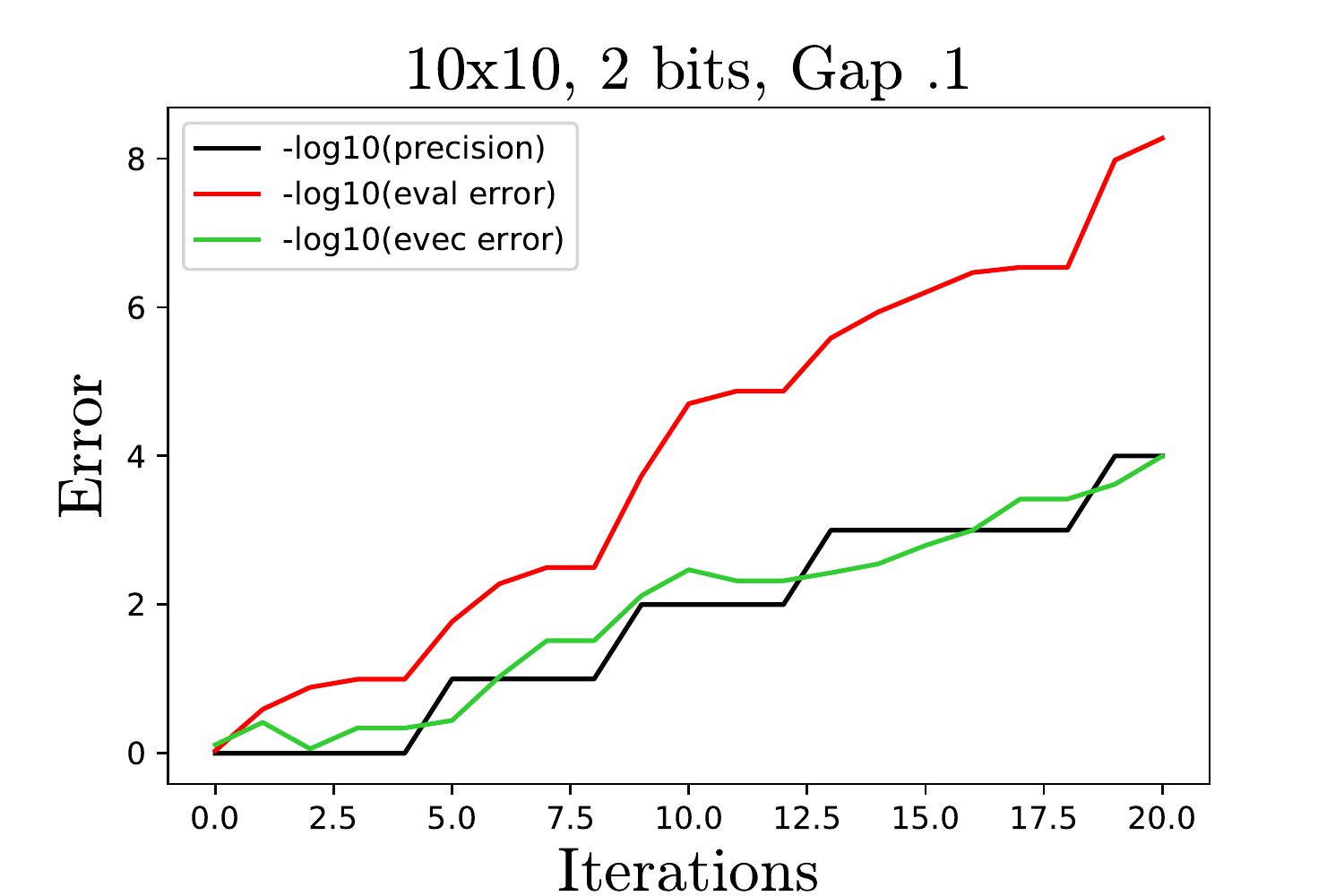}} 
	\subfigure{\includegraphics[width=6cm]{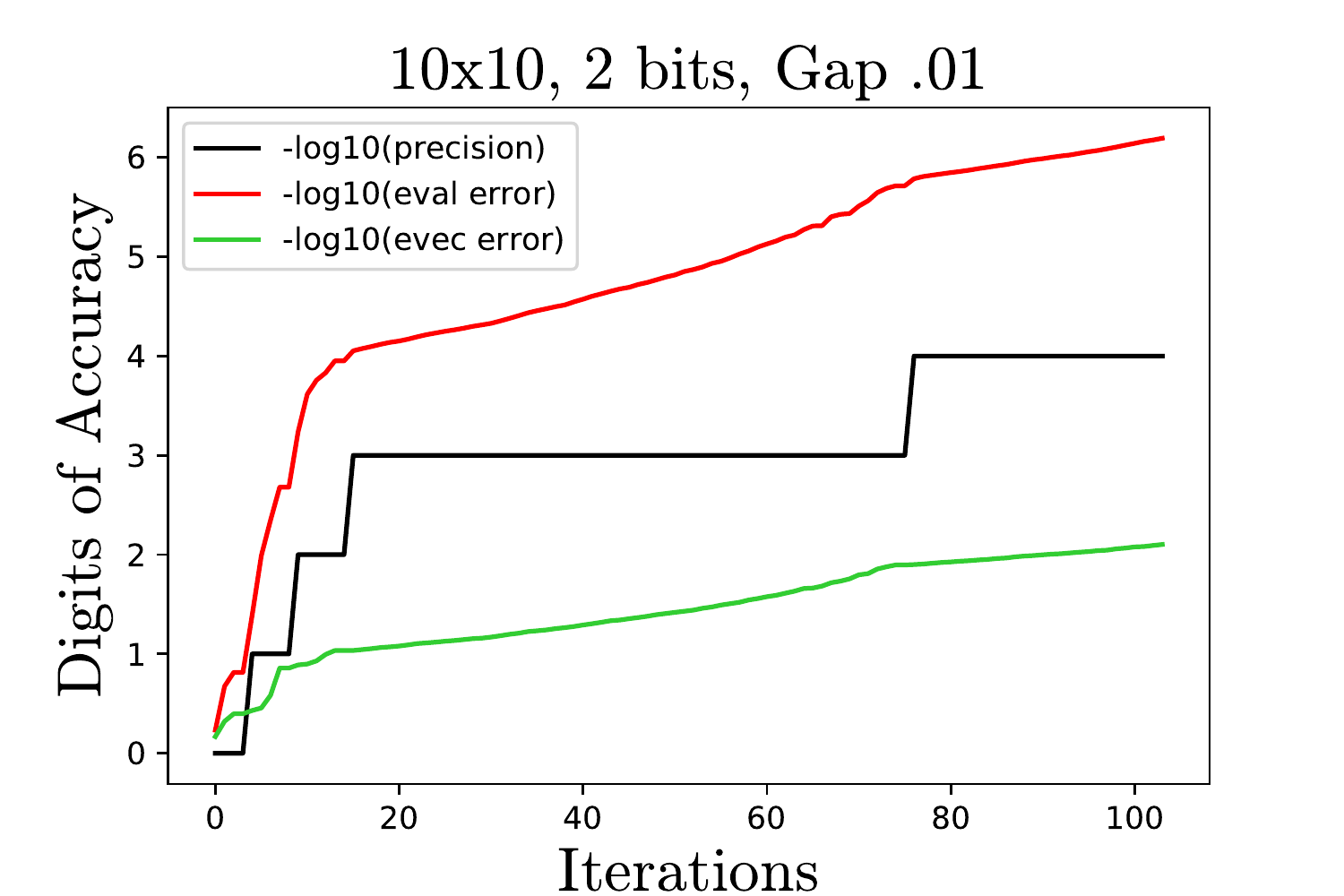}}  
	\caption{Sample plots for two $10 \times 10$ matrices with gap size $.1$ and $.01$. As the gap size decreases, the ratio of eigenvector error to precision and eigenvalue error increases.}
\end{figure}

Figure \ref{fig:gapErrorPlots} has two example error plots demonstrating the slower convergence. Observe that the eigenvector error relative to both the precision and eigenvalue error increases as the gap size decreases.

\indent In the case when there is degeneracy, that is the gap $g$ is  $0$, one might want two eigenvectors that span the eigenspace. This can be accomplished by running the algorithm once to get an approximate eigenvector $\B{v}_1$, replace the matrix $A$ with $A + \alpha \B{v}_1\B{v}_1^t$ where $\alpha > 0$, and run the algorithm again to get the eigenvector $\B{v}_2$. By the spectral theorem for the symmetic matrix $A + \alpha \B{v}_1 \B{v}_1^t$, $\B{v}_1^t\B{v}_2 = 0$ implying that $\B{v}_2$ is an eigenvector for $A$. Replacing $A$ by $A + \alpha \B{v}_1 \B{v}_1^t$ is necessary for numeric purposes. The gap $g$ is never numerically zero, so if the algorithm is run twice on the matrix $A$ even with different randomization, it will often produce the same vector. One can also try to take advantage of the first computation by initializing the approximate eigenvalue to $\lambda_n + \e$ in the second run of the algorithm. The data shown below in Figure \ref{fig:multiplicity} was collected for $\alpha = \frac{tr(A)}{n} - \lambda_n$ and $\e = 1$, and one can see a slight boost in performance in the second run of the algorithm.
\begin{figure}[H]
\label{fig:multiplicity}
	\centering
	\subfigure{\includegraphics[width=6cm]{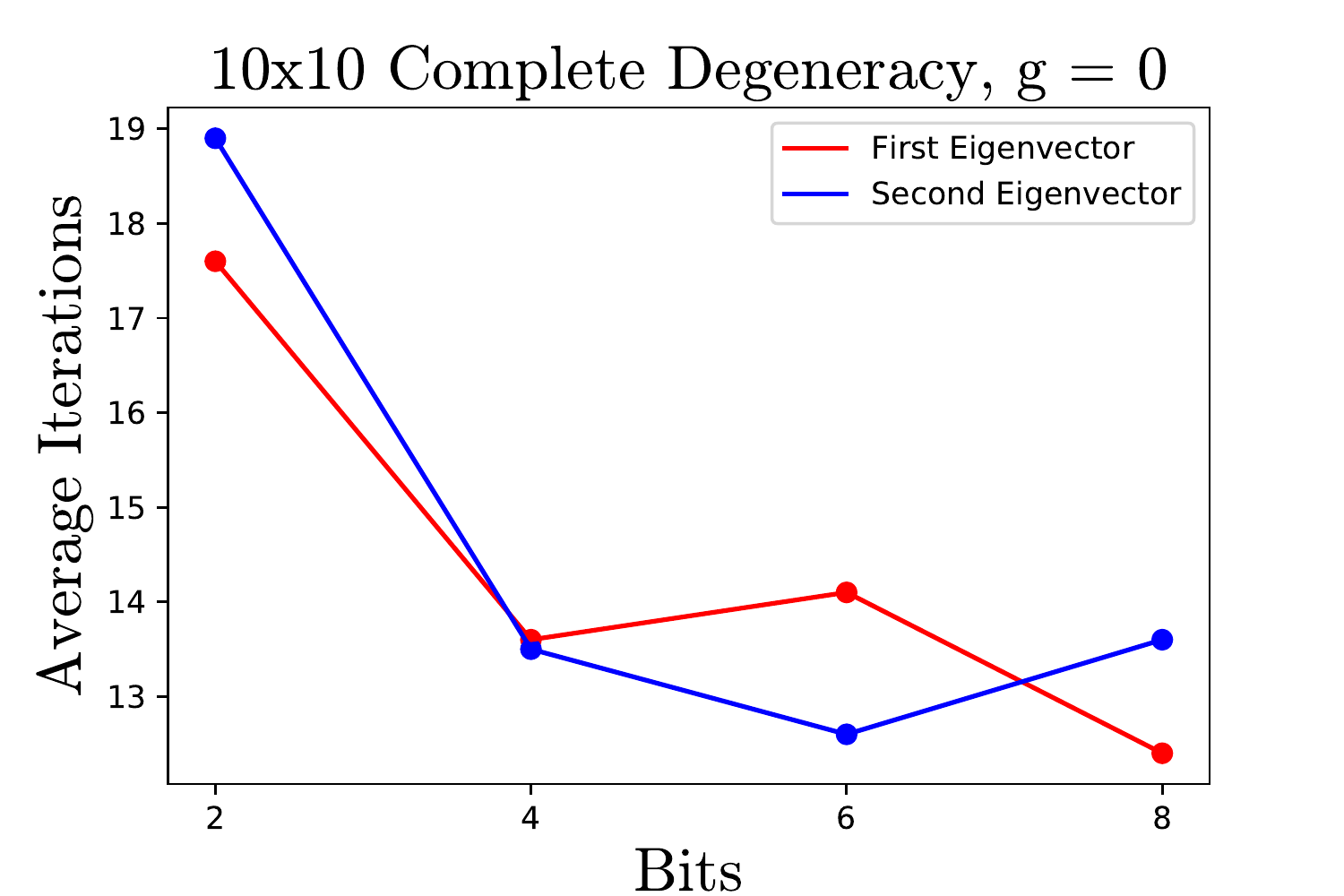}} 
	\subfigure{\includegraphics[width=6cm]{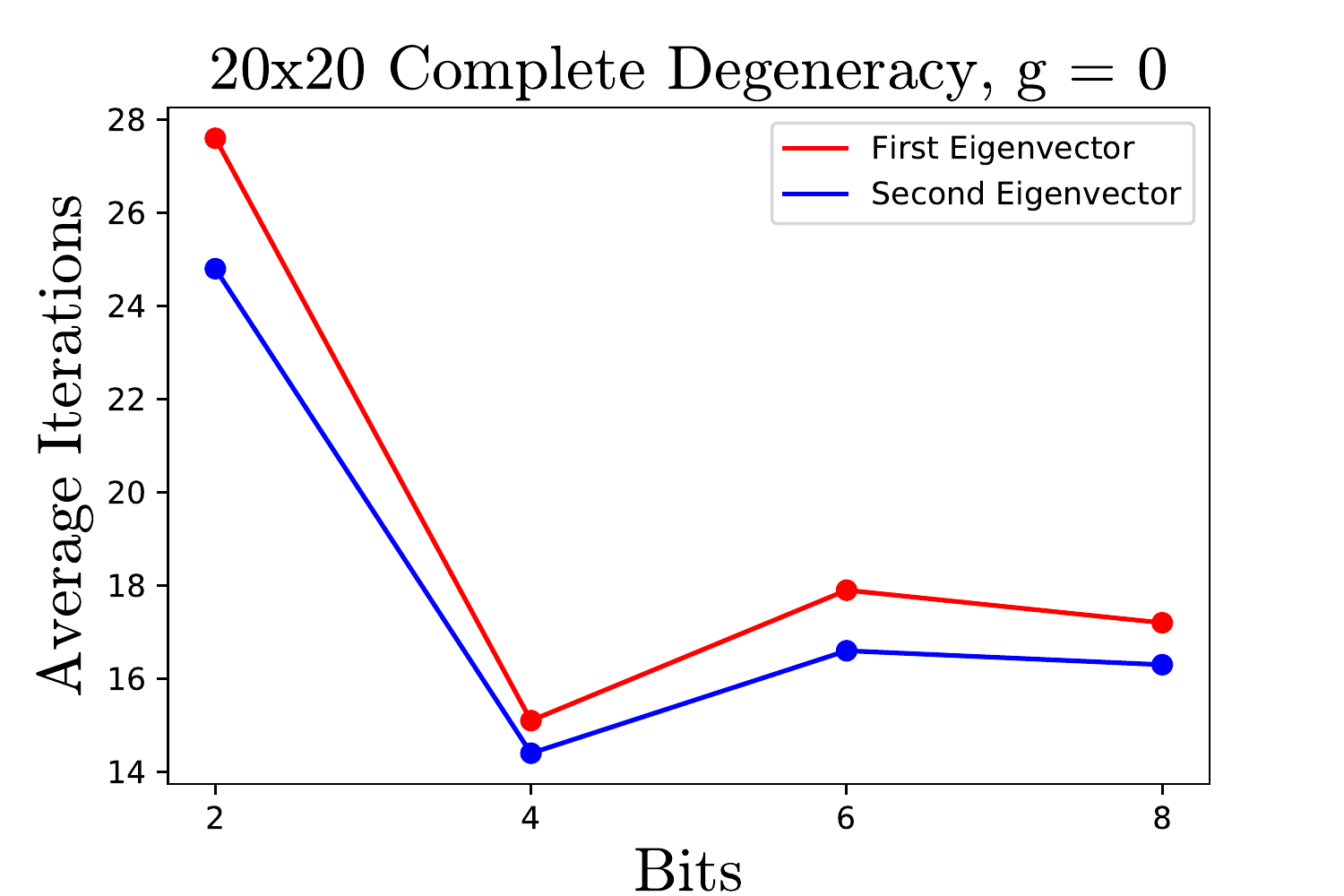}}  
	\caption{Average number of iterations running the algorithm twice on matrices with complete degeneracy. One can see slightly better performance on the second run, particularly on the $20\times20$ matrices}
\end{figure}


\section{Conclusion}
\label{sec:conclusion}
\indent \indent We have proposed and tested an algorithm to find eigenvectors of symmetric matrices by minimizing the corresponding Rayleigh quotient with an iterative steepest-descent method. Initial guesses and subsequent descent directions are found by looking for minima over discretized cubes of various sizes, encoded as QUBO problem which is in turn solved with a SA method. The algorithm is able to reach essentially arbitrary precision even for fairly large matrices.
We have performed a thorough study of the effect of the different parameters, including, the eigenvalue spacing, initial guesses, and number of bits, and the matrix size. We have explored the possibility of using a single bit precision by reformulating the QUBO problem as an Ising problem.
Finally, we have introduced two novel approaches to accelerate the convergence such as biasing and using a larger set of solution from the SA step. These two approaches might be applicable to other QUBO based problems. We encourage the reader to test these algorithms on other annealing devices. 
\bibliographystyle{siamplain}
\bibliography{references}
\end{document}